\pdfoutput=1

\documentclass[sigconf]{acmart}
\setcopyright{none}
\acmConference[]{}{}{}       
\acmPrice{}                   
\acmDOI{}                     
\acmISBN{}                    
\usepackage{siunitx}
\usepackage{amsmath}
\usepackage{graphicx}
\usepackage{dsfont}
\usepackage{booktabs}
\usepackage{array}
\usepackage{booktabs} 
\usepackage{amsfonts}
\usepackage{xspace}
\usepackage{breqn}
\usepackage{hyperref}
\usepackage{caption}
\usepackage{float}
\usepackage{xcolor}

\usepackage{times}
\usepackage{latexsym}
\usepackage{caption}
\usepackage[T1]{fontenc}

\usepackage[utf8]{inputenc}

\usepackage{microtype}

\usepackage{subcaption}

\title{Positional encoding is not the same as context: A study on positional encoding for sequential recommendation} %

\newcommand{\Abs}{\textit{Abs~}}
\newcommand{\AbsCon}{\textit{Abs + Con~}}
\newcommand{\Rotatory}{\textit{Rotatory~}}
\newcommand{\RotatoryCon}{\textit{Rotatory + Con~}}
\newcommand{\LongRotatory}{\textit{Rotatory-Longer~}}
\newcommand{\LongRotatoryCon}{\textit{Rotatory + Con-Longer~}}
\newcommand{\RMHAROPEONE}{\textit{RopeOne~}}
\newcommand{\ROPEMHA}{\textit{RoPE~}}
\newcommand{\RMHA}{\textit{RMHA-4~}}
\newcommand{\LongROPEMHA}{\textit{RoPE-Longer~}}
\newcommand{\LongerRMHA}{\textit{RMHA-4-Longer~}}
\newcommand{\Learnt}{\textit{Learnt~}}
\newcommand{\LearntCon}{\textit{Learnt + Con~}}
\newcommand{\None}{\textit{None~}}
\newcommand{\APE}{\textit{APE~}}
\newcommand{\RPE}{\textit{RPE~}}
\newcommand{\Hit}{\textit{Hit}}
\newcommand{\HitMean}{\textit{Hit Mean}}
\newcommand{\HitDev}{\textit{Hit Dev}}
\newcommand{\AvgDevHit}{\textit{Avg Dev Hit}}

\newcommand{\NDCGMean}{\textit{NDCG Mean}}
\newcommand{\NDCGDev}{\textit{NDCG Dev}}
\newcommand{\AvgDevNDCG}{\textit{Avg Dev NDCG}}

\begin{document}
\author{Alejo López-Ávila}
\affiliation{%
  \institution{Huawei London RC}
  \city{London}
  \country{UK}
}
\email{alejo.lopez.avila@huawei.com}

\author{Jinhua Du}
\affiliation{%
  \institution{Huawei London RC}
  \city{London}
  \country{UK}
}
\email{jinhua.du@huawei-partners.com}

\author{Abbas Shimary}
\affiliation{%
  \institution{Huawei London RC}
  \city{London}
  \country{UK}
}
\email{abbas.shimary@huawei.com}

\author{Ze Li}
\affiliation{%
  \institution{Noah's Ark Lab}
  \city{Hong Kong}
  \country{China}
}
\email{lize23@huawei.com}

\begin{abstract}
The rapid growth of streaming media and e-commerce has driven advancements in recommendation systems, particularly Sequential Recommendation Systems (SRS). These systems employ users' interaction histories to predict future preferences. While recent research has focused on architectural innovations like transformer blocks and feature extraction, positional encodings, crucial for capturing temporal patterns, have received less attention. These encodings are often conflated with contextual, such as the temporal footprint, which previous works tend to treat as interchangeable with positional information. This paper highlights the critical distinction between temporal footprint and positional encodings, demonstrating that the latter offers unique relational cues between items, which the temporal footprint alone cannot provide. Through extensive experimentation on eight Amazon datasets and subsets, we assess the impact of various encodings on performance metrics and training stability. We introduce new positional encodings and investigate integration strategies that improve both metrics and stability, surpassing state-of-the-art results at the time of this work's initial preprint. Importantly, we demonstrate that selecting the appropriate encoding is not only key to better performance but also essential for building robust, reliable SRS models.
\end{abstract}
\maketitle

\section{Introduction}
\textbf{Sequential Recommendation Systems (SRS)} have become integral to personalizing user experiences by predicting the next item of interest based on historical interaction sequences. These systems are widely deployed across e-commerce, social media, and streaming platforms and significantly enhance user engagement and satisfaction. As defined in \cite{DLSRS}, SRS records user-item interactions sequentially, closely related to session-based and session-aware recommendations, which focus specifically on interactions within a single session.

\begin{figure}[t]
    \includegraphics[scale=0.3]{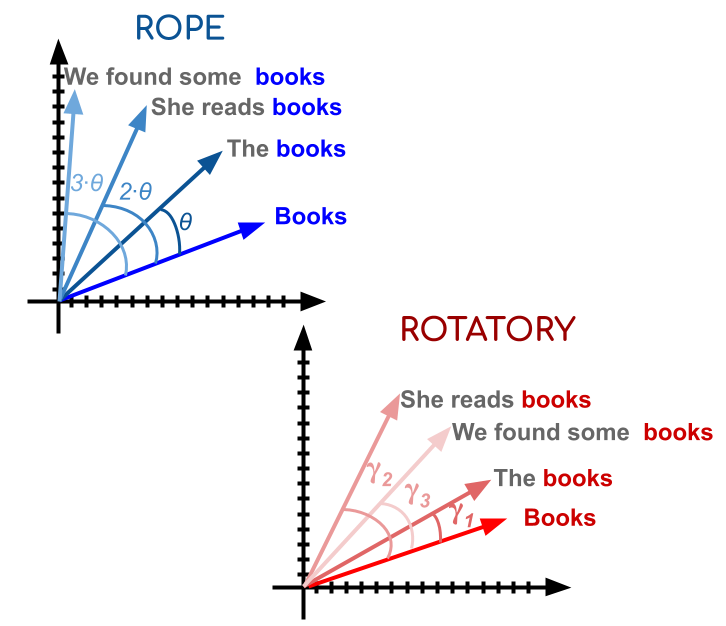}
    \caption{Illustration of our proposed \textbf{\Rotatory} embeddings. Similar to \textbf{\ROPEMHA}, it encodes positional information through the angles of the embeddings. However, unlike \ROPEMHA, which assigns angles proportionally based on position, \Rotatory learns these angles during training. It allows the model to determine the most relevant positional information and adjust their significance accordingly.}
    \label{fig:ROTATORY}
\end{figure}

Capturing the order of data is essential in sequential systems. Initially, Deep Learning techniques utilized Convolutional Neural Networks (CNN) and Recurrent Neural Networks (RNN) to model sequential dependencies. However, \textbf{attentional-based models} have gained prominence due to their superior effectiveness, leading to numerous \textbf{self-attention} approaches \cite{asurveyonsessionbased}. Most state-of-the-art (SOTA) models in SRS now leverage attentional-based networks. A critical component of these systems is \textbf{positional encoding (PE)}, which imparts information about the order of items within a sequence. The choice and implementation of PEs are crucial for optimizing performance and stability.


Despite the widespread adoption of various PE techniques, there is a noticeable gap in research concerning the stability of model performance across different runs. However, many existing publications focus on enhancing performance metrics without adequately addressing the underlying stability issues introduced by different encoding strategies. Other studies underestimate the relevance of positional encoding, considering them part of the context.

To bridge this gap, we aim to investigate the stability problem in SRS models through positional encodings. We conduct extensive experiments to evaluate how different positional encodings influence these models' performance and stability. Building on this foundation, we introduce a novel positional encoding method termed \textbf{\Rotatory}, along with its concatenated variant, \textbf{\RotatoryCon}, and ablations of the previous encoding method.

To summarise, our contribution is fourfold:
\begin{enumerate}
    \item We analyze existing work and highlight that more than the standard usage of $5$ runs is needed to ensure the reliability of results.
    \item We conduct a detailed analysis of various encoding types, their features (such as in-head and affected heads), and perform ablation studies to identify factors that improve stability and performance.
    \item We introduce a novel rotatory-based encoding method, alongside integration techniques, such as concatenation, which have not been explored previously. These contributions aim to improve both performance metrics and training stability.
    \item Our findings show that encoding effectiveness is primarily influenced by dataset deviation, rather than sparsity or data type. This insight, combined with our analysis, offers a clearer guide for using encodings in SRS with transformer architectures.
\end{enumerate}

\section{Related Work}
\begin{table*}[!htp]
    \centering
    \resizebox{0.92\textwidth}{0.1\textheight}{
    \begin{tabular}{|l|c|c|c|c|c|c|}
        \hline
        Encoding & Learnable & Type & Range & Layers & Concatenation & Integration \\
        \hline
        \Abs        & No  & \APE &  Fix. & Vector Encoding   & No  & Add \\
        \AbsCon     & Yes & \APE &  Fix. & Vector Encoding   & Yes & Joint \\
        \Rotatory       & Yes  & \APE &  Unl. & Vector Encoding   & No  & Add \\
        \RotatoryCon    & Yes & \APE &  Unl. & Vector Encoding   & Yes & Joint \\
        \RMHA       & No  & \RPE &  4    & All & No  & Multi. \\
        \ROPEMHA   & No  & \APE &  Unl. & All & No  & Multi. \\
        \RMHAROPEONE   & No  & \APE &  Unl. & First & No  & Multi. \\
        \Learnt     & Yes & \APE &  Fix. & Vector Encoding   & No  & Add \\
        \LearntCon  & Yes & \APE &  Fix. & Vector Encoding   & Yes & Joint \\
        \None       & No  & N/A  &  N/A  & 0   & N/A & N/A \\
        \hline
    \end{tabular}
    }
    \caption{\textbf{Encoding types}: 'Type' specifies absolute or relative. 'Learnable' indicates if encoding parameters are trained, including concatenation for learnable encodings. 'Range' denotes position coverage: 'Fix.' for fixed length, 'Unl.' for unlimited, and $4$ for relative range. 'Layers' refers to encoding application: 'Vector Encoding' adds vectors at the start (left in Fig \ref{fig:APEvsRPE}), 'All' applies to all transformer blocks, and 'First' to the first block. 'Concatenation' shows if concatenation is used. 'Integration' describes the method: 'Add' for vector sum, 'Joint' for concatenation, and 'Multi.' for matrix multiplication in heads or element-wise otherwise. '\None' means no encoding.}
    \label{tab:encodings_table}
\end{table*}

\subsection{Architectures \& Encoding in SRS}

From an architectural perspective, SRS have significantly evolved over the past two decades. Early approaches employed K-Nearest Neighbors (KNN) for item-based recommendations \cite{KNN1, KNN2} and Markov Chains for modeling sequential interactions \cite{MC_order1, MC_orderh1, MC_orderh2}. To address interaction sparsity, matrix factorization methods like DeepFM \cite{DeepFM} and CFM \cite{CFM} were introduced, enhancing the ability to handle sparse data effectively.

With the advent of Deep Learning, RNNs became popular for capturing sequence information and item positions, exemplified by models such as GRU4Rec \cite{GRU4Rec} and DREAM \cite{DREAM}. Concurrently, CNNs were utilized to capture local features and temporal information, as seen in Caesar \cite{Caesar} and NextItNet \cite{NextItNet}, among others \cite{CNN1, CNN2}. \textbf{Attention mechanisms} were first introduced in NARM \cite{NARM}, and transformers with self-attention were pioneered in SRS by AttRec \cite{AttRec}.

Different types of models are used to predict user behavior: \textbf{attribute-aware} \cite{CARCA}, \textbf{context-aware} \cite{S3rec}, and \textbf{time-aware} \cite{TiSASRec}. These models focus on item information, user interaction characteristics, and the timing of interactions, respectively. The integration of transformers has introduced \textbf{positional embeddings}. Some approaches, like CARCA \cite{CARCA}, use time-aware information to replace traditional positional encodings, assuming that the order of items can be deduced from timestamps. However, recent NLP research \cite{asimple} has shown that incorporating positional information directly within the attention mechanism, rather than relying on timestamps, yields better results.

One of the initial models to capture sequential information was GRU4Rec \cite{GRU4Rec}, which extracts data from item sequences. The context-aware DeepFM \cite{DeepFM} combines factorization methods with deep learning to leverage contextual information. Similarly, SASRec \cite{SASRec} employs a self-attention layer to balance short-term intent and long-term preferences, using transformer encoders to compare past behavior with target items via dot-products. This encoder-decoder architecture, which extracts past information while synthesizing target embeddings, led to the development of CARCA \cite{CARCA}.

We focus on positional encoding across various datasets used in these studies. Each work approaches encoding differently: AttRec \cite{AttRec} employs absolute embeddings to incorporate time information into the query and key, while SASRec utilizes absolute learnable position embeddings in the input sequence. CARCA experimented with absolute encodings but observed worse performance compared to using no positional encoding. Recent advancements, such as EulerFormer \cite{EulerFormer}, introduce complex vector attention to integrate semantic and positional information. This variety in encoding strategies highlights the need for a systematic analysis, which we provide in this work.

\label{sec:SRS}
\begin{figure}[h]
    \includegraphics[scale=0.18]{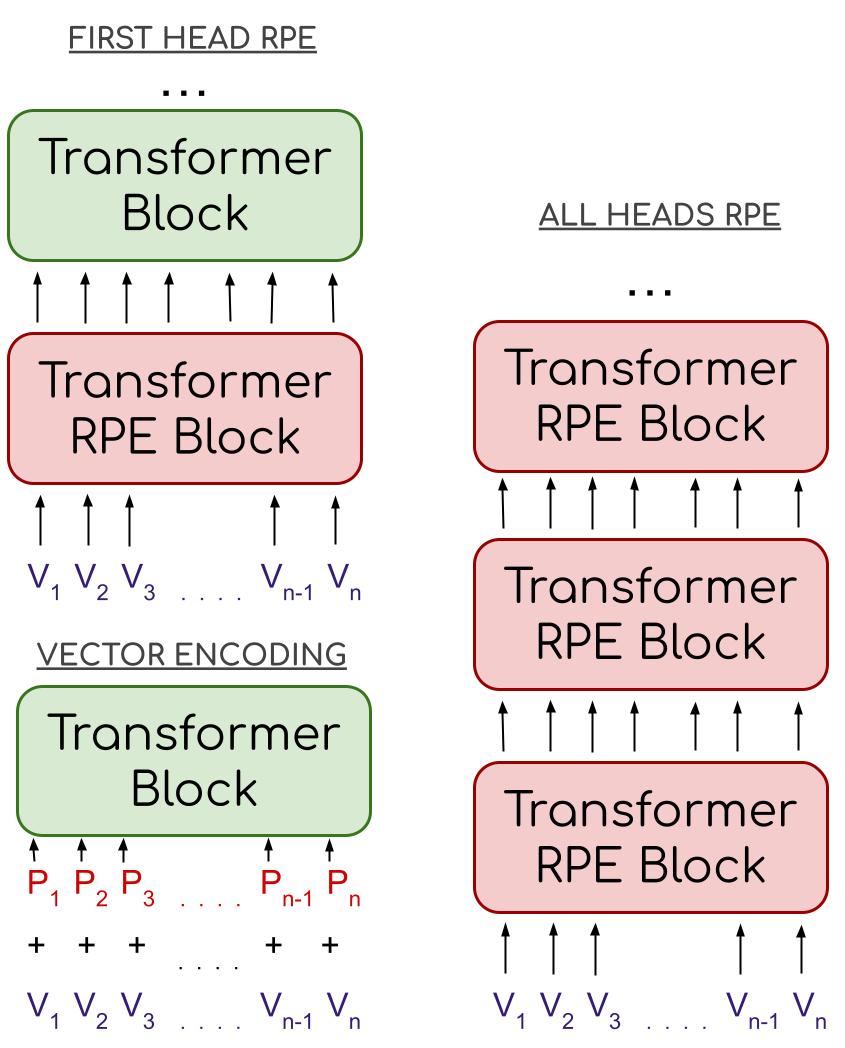}
    \caption{(1) Vector encoding added before Transformer blocks, (2) First head RPE: Relative encoding applied only to the first block (\RMHAROPEONE), and (3) All head RPE: Relative encoding integrated into every block (\RMHA).}
    \label{fig:RPE_heads}
\end{figure}

\begin{figure}[h]
    \includegraphics[scale=0.19]{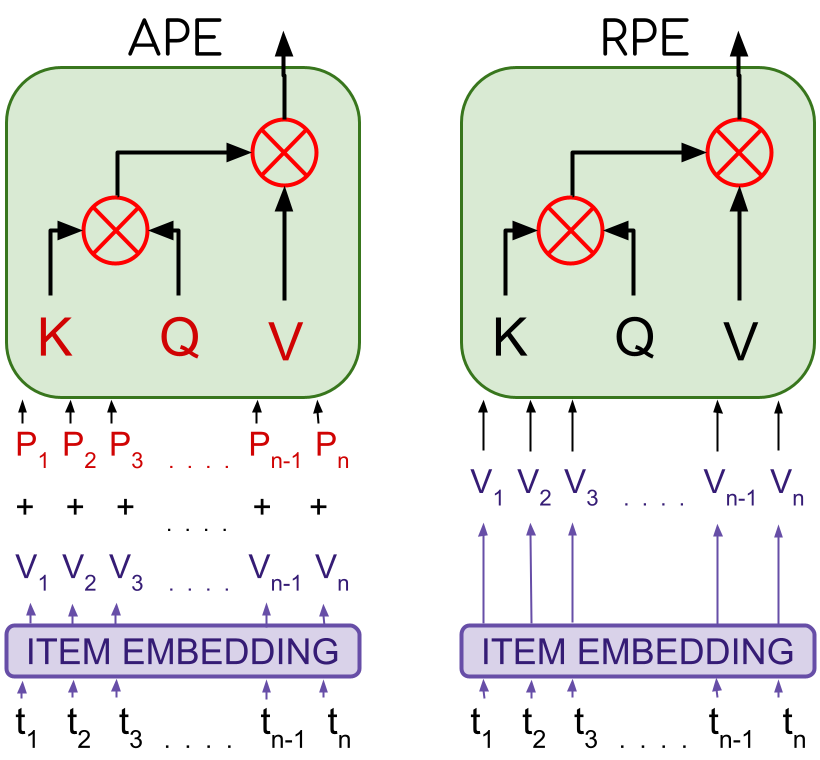}
    \caption{\APE vs \RPE: While \APE encodings introduce the positional information as vectors before the $V$, $Q$ and $K$. \RPE add this information at the coefficient level to $K$ and $Q$. Parts affected by the position information appear in \textcolor{red}{red}.}
    \label{fig:APEvsRPE}
\end{figure}


\subsection{Positional encodings}
\label{sec:PE}

Positional encoding enables transformer-based models to incorporate item order within sequences. Encoding methods can be categorized into absolute and relative encodings \cite{encoding_survey2023} (Figure \ref{fig:APEvsRPE}). In Table \ref{tab:encodings_table}, we have a detailed classification of the main encodings in this paper.

\paragraph{\textbf{Absolute Positional Encoding} (\APE\!)} They assign a unique position vector to each item based on its position in the sequence. These encodings are added before the first Transformer block, providing absolute positional information. Models like SASRec \cite{SASRec} utilize learnable absolute position embeddings, allowing the model to learn position representations during training. There are two common variations of absolute positional encodings: fixed (\Abs) and learnable (\Learnt).

\paragraph{\textbf{Absolute Positional Encoding} (\Abs\!)}
Absolute positional encoding adds fixed values to item embeddings based on their absolute position in the sequence. The most common approach uses sinusoidal functions, as seen in BERT \cite{BERT}. This encoding is added at the input layer as part of the item encoding (Figure, left \ref{fig:APEvsRPE}).

The encoding for an item at position $pos$ in a sequence is calculated using sine and cosine functions:
\begin{equation}
    PE_{(pos, 2i)} = \sin\left(\frac{{pos}}{{10000^{(2i/d)}}}\right)
\end{equation}
\begin{equation}
    PE_{(pos, 2i + 1)} = \cos\left(\frac{{pos}}{{10000^{(2i/d)}}}\right)
\end{equation}
where $i$ is the dimension, and $d$ is the embedding dimension. This encoding is added to the original embedding:
\begin{equation}
    x = x + PE_{(pos, *)}
\end{equation}

\paragraph{\textbf{Learnable Encoding} (\Learnt\!)} 
Learnable position embeddings treat positional information as trainable parameters. Unlike fixed encodings, the model learns optimal position representations based on the data context, denoted as \textbf{learnable encoding} (\Learnt). Positional encodings are added to the input sequence embeddings. 

\paragraph{\textbf{Relative Positional Encoding} (\RPE\!)}
Relative Positional Encoding (Figure \ref{fig:APEvsRPE}, right) encodes the relative distances between items instead of their absolute positions, enabling models to better capture contextual relationships within user interactions. Unlike Absolute Positional Encoding (\APE), which requires fixed input lengths and provides positional information from the start to the end of the sequence, \RPE offers flexibility by focusing on the relationships between items regardless of their absolute positions. A downside of these positional encoding mechanisms is that they are significantly slower during training and inference since they add an extra step in the self-attention layer \cite{DBLP}.

\paragraph{\textbf{Self-attention}}
This mechanism allows models to weigh different parts of the input sequence differently, focusing on relevant information during processing. Introduced in \cite{allyouneed}, the self-attention mechanism computes attention weights based on the similarity between queries (\(Q\)), keys (\(K\)), and values (\(V\)):

\begin{equation}
    \alpha_{ij} = \frac{\exp\left((Q_i K_j^T) / \sqrt{d}\right)}{\sum_{t=1}^{n} \exp\left( (Q_i K_t^T) / \sqrt{d}\right)}
\end{equation}

\begin{equation}
    \text{Attention}(Q, K, V) = \sum_{j=1}^{N} \alpha_{ij} V_j
\end{equation}

\paragraph{\textbf{Relative Positional Encoding Implementation} (\RMHA\!)} This encoding addresses the limitations of \APE by encoding the relative positions within the attention mechanism. Models like NARM \cite{NARM} and TransformerXL \cite{TranformersXL} adopt \RPE to enhance the modeling of contextual dependencies. In \RPE, positional information is incorporated into the key and value projections, allowing the model to focus on the relationships between items dynamically.

The attention weight with \RPE is computed as:

\begin{equation}
    \alpha_{ij} = \frac{\exp\left((Q_i (K_j + a_{ij}^{K})^T) /\sqrt{d}\right)}{\sum_{t=1}^{n} \exp\left((Q_i (K_t + a_{it}^{K})^T) / \sqrt{d}\right)}
\end{equation}

Additionally, positional information can be added to the value projections:

\begin{equation}
    \text{Attention}(Q, K, V) = \sum_{j=1}^{N} \alpha_{ij} (V_j + a_{ij}^{V})
\end{equation}

We will consider the case which bounds the relative indexes to a distance of $4$ and called it \textbf{\RMHA}, Relative Multi-Head Attention with distance $4$.

\paragraph{\textbf{RoPE} (\ROPEMHA\!)}
Rotary Positional Encoding \cite{RoFormer} encodes positional information by applying rotational transformations to the Query and Key matrices. This preserves the dot product between vectors, maintaining effective attention mechanisms. \ROPEMHA is particularly useful for complex item orders, providing a more flexible representation than traditional encodings.

For a two-dimensional vector \(\mathbf{x}\) and angle \(\theta\):
\begin{equation}
    f_{\theta}(\mathbf{x}) = \mathbf{R}(\theta) \mathbf{W} \mathbf{x}
\end{equation}

where \(\mathbf{R}(\theta)\) is a rotation matrix defined by sines and cosines, and \(\mathbf{W}\) represents either \(K\) or \(Q\). Higher-dimensional vectors are split into two-dimensional chunks, with rotations applied to each. To reduce computational costs, RoFormer \cite{RoFormer} implements these rotations using two vector multiplications and one vector addition.


\section{Methodology}
\label{sec:Methodology}

In this section, we present our contributions to positional encodings by introducing variants of existing methods and new encodings.

\subsection{Encoding Variants}

The most commonly used encodings are \Abs encodings, which are vector encodings added to the initial item vector by addition. This approach preserves the original embedding dimension without requiring additional encoding dimensions (Table \ref{tab:encodings_table}). In contrast, non-vector encodings, such as Relative Positional Encoding (RPE) and Rotary Positional Encoding (RoPE), are typically introduced as multiplications within the attention mechanism. These general rules apply depending on whether encodings are relative or absolute. 

To further explore encoding strategies, we introduce an encoding variant that employs concatenation with a linear layer and activation function to revert to the original embedding dimension. To the best of our knowledge, this approach has not been previously utilized in SRS and has seen limited application in NLP. While concatenation benefits include separating semantic and positional information, it introduces a size increase in the architecture, which is particularly critical for Large Language Models. To mitigate this, we incorporate an additional layer, allowing for a gradual introduction of positional information without excessively increasing model complexity.

\paragraph{\textbf{Concatenated Absolute Encoding}}
To enhance positional information, we introduce \textbf{concatenated absolute encoding}, which concatenates the positional encoding to the item embedding:
\begin{equation}
    x = \text{Concat}(x, PE_{(pos, *)})
\end{equation}
A Linear layer is then applied to restore the original embedding dimension without increasing model parameters. This method explicitly incorporates positional information, allowing the model to gradually rely on positional cues and achieving more stable performance (Section \ref{sec:Results}).

\paragraph{\textbf{Concatenated Learnable Encoding} (\LearntCon\!)}
Additionally, we apply a concatenation approach to learnable encodings, denoted as \textbf{concatenated learnable encoding} (\LearntCon\!). However, \LearntCon generally yields worse results compared to the classic residual connection (Section \ref{sec:Results}), likely because concatenation adds complexity in this case, making it harder for the model to effectively learn embeddings. Furthermore, the gradual introduction of positional information is already inherent in the learnable encoding mechanism.

We do not apply concatenated versions for \ROPEMHA and \RMHA, as they are relative encodings.

\paragraph{\textbf{RoPE One Block} (\RMHAROPEONE\!)}
In our experiments (Section \ref{sec:Results}) we found that \RMHA produce more stable results than other encodings. To investigate whether the stability improvements of \RMHA were due to introducing positional information at each Transformer block or its role as an \RPE-based head, we tested \ROPEMHA. The results showed that that \ROPEMHA was in a middle point between relative and absolute encodings regarding stability. 

We hypothesize that this intermediate stability is due to the application of \ROPEMHA across all Transformer blocks (($3$) in Figure \ref{fig:RPE_heads}). To investigate this further, we introduced \textbf{\RMHAROPEONE}, which applies \ROPEMHA exclusively in the first Transformer block (($2$) in Figure \ref{fig:RPE_heads}). As detailed in Section \ref{sec:Results}, this modification yields slightly improved results. This implies that the rotary aspect, rather than merely being relative, is key to the observed improvements and leads us to the new encoding introduced below.

\subsection{Rotatory Encoding}
We propose a new encoding method termed \Rotatory (Figure \ref{fig:ROTATORY}), with its concatenated variant referred to as \RotatoryCon This approach applies a rotation to define the position of each item. Unlike traditional methods that add positional information during self-attention, our method incorporates these rotations at the initial item embeddings, similar to sinusoidal encoding (\Abs\!). Vector embeddings offer the advantage of faster training and inference \cite{DBLP}. Therefore, we adopt a formulation similar to \ROPEMHA, applying the rotations at the outset. We enhance flexibility by making the rotation angles learnable, allowing the model to capture both absolute and relative positional relationships effectively.

The positional encoding for an item at position $pos$ and embedding dimension $i$ is calculated using a trigonometric approach, modified to include alternating sine signs for a rotational effect. The encoding is defined as follows:

\begin{equation}
    PE_{(pos, 2i)} = (-1)^i \sin\left( \theta_{(pos, i)} \right)
\end{equation}
\begin{equation}
    PE_{(pos, 2i + 1)} = \cos\left( \theta_{(pos, i)} \right)
\end{equation}

Here, $pos$ is the position of the item in the sequence, $i$ represents the embedding dimension, and $\theta_{(pos, i)}$ is a learnable angle defined as:

\begin{equation}
    \theta_{(pos, i)} = \frac{E_{pos}}{10000^{(2i/d)}} \cdot 2\pi , \quad \forall pos \in [0, L), \forall i \in [0, H)
\end{equation}

where $E_{pos} \in \mathbb{R}^{L \times H}$ are learnable embeddings. The term $(-1)^i$ in the sine function alternates the sign, introducing the rotation effect in the encoding. These positional encodings are then added to the original item embeddings by addition:

\begin{equation}
    x'_{(pos)} = x_{(pos)} + PE_{(pos)}
\end{equation}

where $x_{(pos)}$ is the original semantic embedding of the item at position $pos$, and $PE_{(pos)}$ is the learnable \textbf{\Rotatory} encoding. In the case of \textbf{\RotatoryCon}, the positional encoding is concatenated with the original item embedding and passed through a learned linear transformation for further adaptation. This learnability introduces flexibility, enabling the model to capture both absolute and relative positional information in a dynamic manner, which leads to improved performance in handling complex sequence relationships.


\section{Experimental Setup}

\subsection{Datasets}
\label{sec:datasets}
We selected several Amazon datasets to evaluate performance of various encoding types. For this purpose, we utilised four distinct real-world datasets extracted from product reviews on Amazon.com. These datasets are widely used for SRS under the leave-one-out protocol \citep{data1, data3, data2, data5, data4, S3rec}. Since multiple versions of these datasets exist, we opted for the preprocessed versions provided by \cite{CARCA}.

\begin{table}[htbp]
    \centering
    \captionsetup{justification=centering} 
    \resizebox{0.48\textwidth}{0.055\textheight}{
        \begin{tabular}{lcccc}
            \toprule
            \textbf{Dataset} & \textbf{Users} & \textbf{Items} & \textbf{Interactions} & \textbf{Attributes} \\
            \midrule
            Men & 34,244 & 110,636 & 254,870 & 2,048 \\
            Fashion & 45,184 & 166,270 & 358,003 & 2,048 \\
            Games & 31,013 & 23,715 & 287,107 & 506 \\
            Beauty & 52,204 & 57,289 & 394,908 & 6,507 \\
            \bottomrule
        \end{tabular}
    }
    \caption{Dataset Statistics.}
    \label{tab:Interactions}
\end{table}
\vspace{-14pt}
\begin{table}[htbp]
    \centering
    \captionsetup{justification=centering} 
    \resizebox{0.48\textwidth}{0.055\textheight}{
        \begin{tabular}{lcccc}
            \toprule
            \textbf{Dataset} & \textbf{\AvgDevHit} & \textbf{\AvgDevNDCG} & \textbf{CI-length} & \textbf{Density} \\
            \midrule
            Men   & 3.9143 & 8.2742 & 5.6686 & $6.727 \cdot 10^{-5}$\\
            Fashion & 4.87 & 9.7814 & 9.2257 & $4.765 \cdot 10^{-5}$ \\
            Games  & 1.124 & 0.952  &1.616   & $3.904 \cdot 10^{-4}$ \\
            Beauty & 0.935 & 0.87   & 1.22   & $1.320 \cdot 10^{-4}$ \\
            \bottomrule
        \end{tabular}
    }
    \caption{Deviations among the different datasets without Encoding. This pattern is generally preserved along the different encodings. The density refers to the inverse of Sparsity.}
    \label{tab:deviations}
\end{table}

\vspace{-10pt}

Covering various product categories, these datasets offer a broad source for training and evaluating recommendation models. They include essential features for each interaction, such as user IDs, item IDs, timestamps, and additional vector-based contextual information. As shown in Tables \ref{tab:Interactions} and \ref{tab:deviations}, they exhibit diversity in terms of sparsity, defined as the ratio of missing user-item interactions. This will be a crucial aspect of our analysis, along with the discreteness of the features.

\paragraph{1.Beauty} \cite{data5, S3rec, data3} The dataset includes discrete and categorical attributes for all beauty products, including fine-grained categories and brands. Mainly categorical and discrete features.

\paragraph{2.Video Games} \cite{data5, data3} The Video Games sub-dataset includes user interactions, reviews, and product details specific to the video game category, such as price, brand and categorical features. Most of the attributes here are discrete and categorical.

\paragraph{3.Men} \cite{data1} The men's dataset encompasses a comprehensive collection of items falling under men's clothing. The attributes are dense vectors from image-based features extracted from the last layer of a ResNet50 \cite{ResNet50} on the ImageNet dataset \cite{ImageNet}.

\paragraph{4.Fashion} \cite{data1, data2} It contains six categories for men's and women's clothing. The features were extracted as dense using the same ResNet50 approach.

\subsection{Models and Training}
To implement CARCA, we converted the original TensorFlow model into PyTorch\footnote{https://github.com/researcher1741/Position_encoding_SRS}. We used the loss implementation from \cite{rapso}. The original TensorFlow version built upon prior models within the same framework. During our evaluation, we observed that the implementations checked metrics every $20$ epochs, assuming that the optimal performance point would be reached later. This approach was likely adopted due to the extensive number of epochs, exceeding $1000$, which increases the computational cost of evaluations. To achieve more reliable evaluations, we implemented a logarithmic measurement of the metrics, which resulted in slightly improved outcomes in certain cases.

As in previous works \cite{CARCA, SASRec}, we use an ADAM optimiser to minimise the binary cross-entropy loss of the CARCA model while masking the padded items to prevent them from contributing to the loss function. For a given sequence of items interactions for a user \(u\) as $\{ i^{u}_{1}, \ldots, i^{u}_{N} \}$, we create an input list as $I^{u+} = \{ i^{u}_{1}, \ldots, i^{u}_{N-1} \}$ by deleting the last item of the list, a positive target list as $T^{u+} = \{ i^{u}_{2}, \ldots, i^{u}_{N} \}$ by shifting the input list by one, and a negative target list as $T^{u-} = \{ i^{rand_1}, \ldots, i^{rand_{N-1}} \}$ generated by random negative items. Then the loss is given by:

\small
\begin{equation*}
- \sum_{u \in U} \sum_{r \in T^{u+} \cup T^{u-}} \left[ Y_r \log(\hat{Y}_r) + (1 - Y_r) \log(1 - \hat{Y}_r) \right]
\end{equation*}
\normalsize
Here, \(U\) represents the set of users, \(T^{u+} \cup T^{u-}\) denotes the union of positive and negative items for user \(u\), \(Y_r\) is the observed interaction label for item \(r\), and \(\hat{Y}_r\) is the predicted probability of interaction with item \(r\).

\begin{table}[htbp]
    \centering
    \captionsetup{justification=centering} 
    \resizebox{0.48\textwidth}{!}{
\begin{tabular}{l|rrrr}
\toprule
\textbf{Encoding} & \textbf{\AvgDevHit} &  \textbf{\AvgDevNDCG} & \textbf{runs} &  \textbf{CI-length} \\ 
\midrule
\Abs   & 5.43 & 8.25 &  9.38 &  6.71 \\
\AbsCon & 3.52 & 5.56 &  7.12 &  5.35 \\
\Learnt & 4.79 & 7.64 &  7.88 &  7.05 \\
\LearntCon  & 3.17 & 6.36 &  6.62 &  4.42 \\
\None  &2.35 & 5.26 &  8.00 &  3.29 \\
\RMHA  &0.51 & 1.08 &  6.00 &  0.88 \\
\ROPEMHA    &2.38 & 5.45 &  5.00 &  4.04 \\
\Rotatory   &2.46 & 5.01 &  8.38 &  3.48 \\
\RotatoryCon &2.48 & 5.18 &  6.75 &  3.49 \\
\midrule
\RMHAROPEONE &2.82 & 6.54 &  4.29 &  5.46 \\
\LongROPEMHA & 3.33 & 8.15 &  5.25 &  5.99 \\
\LongRotatory    &1.05 & 1.04 &  4.67 &  2.08 \\
\LongRotatoryCon &0.99 & 1.24 &  6.00 &  1.59 \\
\LongerRMHA &0.71 & 4.00 &  6.00 &  1.14 \\
\bottomrule
\end{tabular}
    }
    \caption{Deviations among the different encodings for the default upper bound values ($0.0001$ for all except NaN for Games). Relative encoding in all the heads, \RMHA, presents more stability, followed by our \Rotatory versions.}
    \label{tab:deviations_encoding}
    \vspace{-10pt}
\end{table}

\begin{table*}[ht]
    \centering
    \captionsetup{justification=centering} 
    \resizebox{0.85\textwidth}{0.16\textheight}{
    \begin{tabular}{lllccccc}
        \toprule
        Dataset & Act & encoding &  nmax & \HitMean & \HitDev & \NDCGMean &  \NDCGDev \\
        \midrule
         Beauty & silu & \LongRotatory & 0.0001 & 61.87 & 1.04 & 42.60 & 0.98 \\
         Beauty & silu &     \Rotatory & 0.0001 & 61.72 & 1.14 & 42.54 & 1.52 \\
         Beauty & silu &  \RotatoryCon & 0.0001 & 61.68 & 1.04 & 42.31 & 0.97 \\
         Beauty & leaky &  \RotatoryCon & 0.0001 & 61.16 & 0.77 & 42.83 & 0.48 \\
        \midrule
         Men &  silu & \LongerRMHA & 0.0001 & 70.13 & 0.42 & 46.41 & 4.18 \\
         Men &  silu &   \RMHA     & 0.0001 & 69.70 & 0.64 & 43.75 & 1.44 \\
         Men & leaky &   \None     & 0.1000 & 69.69 & 1.28 & 57.94 & 1.69 \\
         Men & leaky & \LongerRMHA & 0.0001 & 68.72 & 1.47 & 43.46 & 0.79 \\
        \midrule
         Fashion & silu  & \RMHA       & 0.0001 & 77.26 & 0.24 & 49.75 & 2.33 \\
         Fashion & silu  & \LongerRMHA & 0.0001 & 77.00 & 0.48 & 49.40 & 0.54 \\
         Fashion & leaky & \RMHA      & 0.0001 & 76.46 & 0.25 & 49.49 & 2.18 \\
         Fashion & leaky & \LongerRMHA& 0.0001 & 76.20 & 0.51 & 49.85 & 3.92 \\
        \midrule
         Games & leaky & \RotatoryCon     & NaN & 80.62 & 0.55 & 56.07 & 0.79 \\
         Games & silu  & \LongRotatoryCon & NaN & 80.29 & 0.47 & 55.18 & 0.55 \\
         Games & leaky & \LongRotatoryCon & NaN & 80.21 & 1.51 & 55.06 & 1.93 \\
         Games & silu  & \Learnt          & NaN & 79.82 & 1.88 & 54.30 & 2.95 \\
        \bottomrule
    \end{tabular}
    }
    \caption{The table displays the top $4$ results for each dataset on cases with a standard deviation for \HitMean below $3$.}
    \label{tab:stable_best}
\end{table*}

\begin{table*}[ht]
    \centering
    \captionsetup{justification=centering} 
    \resizebox{0.85\textwidth}{0.16\textheight}{
    \begin{tabular}{lllccccc}
        \toprule
        Dataset & Act &   encoding &  nmax & \HitMean & \HitDev & \NDCGMean &  \NDCGDev \\
        \midrule
        Beauty & leaky& \AbsCon & 0.0001 & 67.93 & 4.17 & 48.71 & 3.59 \\
        Beauty & silu & \Learnt & 0.0001 & 67.62 & 6.95 & 46.33 & 7.08 \\
        Beauty & silu & \Abs & 0.0001 & 67.27 & 10.59 & 45.34 & 10.75 \\
        Beauty & silu & \AbsCon & 0.0001 & 65.87 & 5.23 & 45.61 & 5.12 \\
        \midrule
        Men & leaky & \LearntCon & 0.1000 & 73.86 & 7.18 & 58.89 & 8.15 \\
        Men & leaky & \Learnt & 0.0001 & 72.08 & 4.44 & 56.55 & 9.98 \\
        Men & leaky & \LearntCon & 0.0001 & 71.34 & 7.13 & 57.92 & 13.07 \\
        Men & leaky & \RotatoryCon & 0.1000 & 70.92 & 3.92 & 59.84 & 4.88 \\
        \midrule
        Fashion & silu & \RMHA & 0.0001 & 77.26 & 0.24 & 49.75 & 2.33 \\
        Fashion & silu & \LongerRMHA & 0.0001 & 77.00 & 0.48 & 49.40 & 0.54 \\
        Fashion & leaky & \RMHA & 0.0001 & 76.46 & 0.25 & 49.49 & 2.18 \\
        Fashion & leaky & \LongerRMHA & 0.0001 & 76.20 & 0.51 & 49.85 & 3.92 \\
        \midrule
        Games & leaky & \RotatoryCon & NaN & 80.62 & 0.55 & 56.07 & 0.79 \\
        Games & silu & \LongRotatoryCon & NaN & 80.29 & 0.47 & 55.18 & 0.55 \\
        Games & leaky & \LongRotatoryCon & NaN & 80.21 & 1.51 & 55.06 & 1.93 \\
        Games & silu & \Learnt & NaN & 79.82 & 1.88 & 54.30 & 2.95 \\
        \bottomrule
        \end{tabular}
        }
    \caption{The table displays the top $4$ results for each dataset on cases with a standard deviation for \HitMean below $12$.}
    \label{tab:General}
    \vspace{-10pt}
\end{table*}

\subsection{Metrics}
We employ two widely used metrics for Next-Item Recommendation and SRS: Hit@10 and Normalised Discounted Cumulative Gain (NDCG).

\textbf{Hit@10}: This metric measures the fraction of times the ground truth next item appears within the top $10$ recommended items. For simplicity, we refer to Hit@10 as \Hit, its mean as \HitMean, its devisation as \HitDev~ and the average of the deviations as \AvgDevHit.

\textbf{NDCG}: Normalised Discounted Cumulative Gain evaluates the ranking quality by considering both the relevance of items and their positions in the recommended list. It is defined as:

\begin{equation}
    NDCG = \frac{DCG}{IDCG}
\end{equation}

where $DCG$ (Discounted Cumulative Gain) is calculated as the sum of the relevance scores of items at each position in the ranked list, discounted by their position:

\begin{equation}
    DCG = \sum_{i=1}^{n} \frac{2^{rel_i} - 1}{\log_2(i+1)}
\end{equation}

and $IDCG$ (Ideal Discounted Cumulative Gain) represents the maximum possible $DCG$ for a perfectly ranked list:

\begin{equation}
    IDCG = \sum_{i=1}^{n} \frac{2^{max(rel_i, 0)} - 1}{\log_2(i+1)}
\end{equation}

Here, $rel_i$ denotes the relevance score of the item at position $i$ in the ranking. We will denote its mean as \NDCGMean, its deviation as \NDCGDev~ and the average of the deviations as \AvgDevNDCG.


\section{Results}
\label{sec:Results}

We present comprehensive results for all datasets and encodings in Appendix \ref{sec:complete_tables}. Due to space limitations, only selected sub-tables are included in this section. Furthermore, Appendix \ref{sec:SASRec} presents an additional experiment using SASRec++, confirming the consistency of our results. Details about the resources used for these experiments are provided in Appendix \ref{sec:resources}.

In addition to positional encoding, we analyze factors, such as maximum values and activation functions, that may influence model stability. In CARCA \cite{CARCA}, an upper bound of $0.0001$ ($nmax$ in our Tables) for the encoding vectors was identified as optimal across all datasets except for Games. To investigate whether the optimal upper bound varies with different encodings, we experimented with maximum values of $0.1$ and $0.0001$. Generally, an upper bound of $0.0001$ yielded better results, with a few exceptions. Furthermore, we compared activation functions by replacing $leakyrelu$ with $silu$. The $silu$ activation function resulted in slight improvements in stability; however, the correlation was not strong.

\subsection{Stability}
\label{sec:Stability}
A common practice in the literature is to present the average results of $5$ runs with different random seeds. However, our preliminar experiments revealed significant variability between individual runs. 
To better understand this inconsistency, we analyzed the standard deviation across different trainings. Detailed deviation and mean values are provided in all the training tables. Additionally, we computed the $95\%$ confidence intervals for these metrics and included them in the complete results tables in Appendix \ref{sec:complete_tables}.

\paragraph{Inter-Dataset Deviation:}
The stability of results varies across different datasets. Beauty is the most stable, showing an average Hit deviation of $0.9$ and an average NDCG deviation of $0.87$, while Fashion has the highest deviations. Datasets with lower sparsity, such as Beauty and Games, generally exhibit greater stability. In contrast, datasets with higher sparsity, like Men and Fashion, show reduced stability. Additionally, there is a correlation with data type: lower-sparsity datasets often use discrete embedding representations. As discussed in Section \ref{sec:sparsity_reduction}, this stability pattern persists even when we adjust sparsity levels. By reducing datasets to achieve comparable sparsity, we show that both encoding selection and stability challenges remain for each data type.

\paragraph{Inter-Encoding Deviation:} Tables \ref{tab:deviations_encoding} and \ref{tab:deviations_encoding_all} (Appendix) reveal that some encodings yield more stable results than others. In particular, the concatenated versions of \Abs, \Learnt, and \Rotatory demonstrate greater stability (Table \ref{tab:deviations_encoding_all}). Under the most stable scenario with an upper bound of $0.0001$, as shown in Table \ref{tab:deviations_encoding}, \AbsCon and \LearntCon remain significantly more stable. However, \RotatoryCon shows no significant difference here, presenting deviations similar to \None, likely due to the need to learn the angles. 

The most notable encoding for stability is \RMHA\!, which, combined with the $0.0001$ upper bound, produces the most stable results, with an \AvgDevHit~ of $0.51$ compared to $2.36$ for the no-encoding case. This pattern persists even with higher upper bounds, suggesting that \RMHA offers excellent stability, which can also be observed in the loss graph, as detailed in Appendix \ref{sec:losses}. As we will discuss in Section \ref{sec:sparsity_reduction}, this encoding is particularly worth considering in unstable scenarios.

\paragraph{Encoding Ablation:} To assess whether the stability is due to \RMHA~'s positional information at each transformer block or its \RPE~-based nature, we tested \ROPEMHA, an \APE\!. The results rank \ROPEMHA between \RMHA and \Learnt/\Abs\!, showing no significant change in stability compared to \None (Table \ref{tab:deviations_encoding}). We hypothesized that this lack of change in stability could be due to \ROPEMHA\!'s application across all transformer blocks. To test this, we introduced \RMHAROPEONE~, which applies \ROPEMHA only in the first block. Results in Table \ref{tab:deviations_encoding_all} show that \RMHAROPEONE slightly outperforms \ROPEMHA\!, while Table \ref{tab:deviations_encoding} shows similar performance with a slight lower stability. This suggests that positional information at each transformer block does not significantly impact stability. Finally, we introduced \Rotatory, which incorporates rotatory properties in a vector encoding. The results showed similar stability to \ROPEMHA and its ablation \RMHAROPEONE~. This led us to the conclusion that the rotatory nature of some encodings and the relative encoding nature of \RMHA contribute to the stability of the experiments.

\subsection{Best Results}
\label{sec:Best}
\paragraph{Which is the best encoding?} The optimal encoding depends on the maximum deviation considered. For instance, examining the Beauty dataset in Table \ref{tab:beauty}, we observe lower deviations. However, in cases like \Learnt with $nmax = 0.1$ and leakyrelu, despite achieving $74$ in \HitMean, the \HitDev~ increases to $17.19$. This highlights the trade-off between performance and stability, as some encodings may offer better performance metrics at the cost of higher instability. These findings support the conclusions of \cite{seed}, which suggest that, in certain scenarios, selecting a robust set of random seeds can be more important than choosing the best encoding method.

\paragraph{Maximum deviation of $12$:} When setting the maximum \HitDev~ to $12$, we get Table \ref{tab:General}. In the CARCA \cite{CARCA}, an ablation analysis of the Men dataset revealed that using \Abs results in poorer performance than \None\!. Similar results were observed for the Men in our experiments, where both cases showed a deviation of around $6$, suggesting that the observed effects were more due to initialization than encoding choice. With a maximum of $12$ on \HitDev, Men and Beauty achieve their best performance with \Learnt\!-like and \Abs\!-like encodings, respectively, although both are unstable. In contrast, Fashion and Games reach their optimal results with \RMHA and \Rotatory\!, respectively, both of which are stable.

\begin{table*}[ht]
    \centering
    \captionsetup{justification=centering} 
    \resizebox{0.85\textwidth}{0.11\textheight}{
    {
    \begin{tabular}{lllccccc}
        \toprule
        Dataset & Act &   encoding &  nmax & \HitMean & \HitDev & \NDCGMean &  \NDCGDev \\
        \midrule
        Submen &  leaky &  \RMHA & 0.1000 & 77.28 & 2.66 & 68.45 & 3.65 \\
        Submen &  leaky &  \RMHA & 0.0001 & 75.96 & 2.71 & 66.03 & 4.00 \\
        \midrule
        Subfashion & leaky & \RMHA & 0.0001 & 79.72 & 2.85 & 71.57 & 3.66 \\
        Subfashion & leaky & \None & 0.1000 & 32.94 & 1.89 & 16.77 & 3.00 \\
        \midrule
        Subgames & leaky & \RotatoryCon & NaN & 52.23 & 1.57 & 31.51 & 0.96 \\
        Subgames & silu  & \RotatoryCon & NaN & 50.73 & 0.89 & 30.48 & 0.72 \\
        Subgames & leaky &        \RMHA & NaN & 50.43 & 0.95 & 29.96 & 0.87 \\
        Subgames & leaky &        \None & NaN & 49.53 & 1.01 & 28.84 & 0.69 \\
        \midrule
        Submen3 & leaky & \RotatoryCon & 0.1000 & 73.12 & 2.38 & 62.86 & 2.99 \\
        Submen3 & silu  & \RotatoryCon & 0.0001 & 72.77 & 2.96 & 62.56 & 4.31 \\
        \bottomrule
    \end{tabular}}
    }
    \captionsetup{width=.82\textwidth}
    \caption{The table displays the top $4$ results for each ablation dataset on cases with a \HitDev below $3$. In the case of the Fashion dataset ablation we did not get any training below $3$ in the deviation.}
    \label{tab:stable_best_ablation}
    \vspace{-10pt} 
\end{table*}

\paragraph{Maximum deviation of $3$:} Reducing the maximum \HitDev~ to $3$ yields optimal results, as presented in Table \ref{tab:stable_best}. In this Table, the best encodings are \RMHA and \Rotatory (or its concatenated version), which correlate strongly with sparsity, datatype, and stability (low-deviation). As discussed in Section \ref{sec:sparsity_reduction}, datasets with high-deviation without encoding benefit from the stability provided by \RMHA, while low-deviation datasets like Beauty and Video Games perform best with the purely rotatory approach of \Rotatory. Even among stable datasets, these results surpass the previous SOTA benchmarks (CARCA) at the time. We also observe a significant improvement over the \None case for high-deviation datasets, with gains exceeding $5\%$, while achieving greater stability.

\paragraph{Impact of Extended Training on Encoding Stability:} Ablation studies were conducted on the encoding techniques to investigate the source of stability. Our initial experiments used hyperparameters similar to those of the original CARCA model (Appendix \ref{sec:Hyperparameters}). Upon reviewing the loss metrics, we found that several top-performing models did not converge, as both loss and performance metrics continued to improve. To address this, we extended the training by an additional $400$ epochs beyond the default values specified in Table \ref{tab:hyperparameters}. Despite this extension, the improvements in model performance were minimal, indicating that the optimizer may have been oscillating around a local minimum.


\subsection{Sparsity and Deviation Reduction}
\label{sec:sparsity_reduction}

As seen in Section \ref{sec:Best}, two encodings, \RMHA and \RotatoryCon, consistently provide the best results (with maximum \HitDev~ of $3$). There appears to be a correlation between the deviation of datasets and the choice of optimal encoding. For high-deviation datasets like Men and Fashion, \RMHA yields better results, whereas for low-deviation datasets like Games and Beauty, \RotatoryCon performs best. Importantly, both Men and Fashion share two other key characteristics: dense vector representations and higher-sparsity. This makes it difficult to attribute the performance solely to the default dataset deviation. The complete tables for the subset results are available in Appendix \ref{sec:complete_tables_deviation}.

\paragraph{Sparsity Reduction:} 
To investigate this further, We conducted a sparsity reduction experiment by creating subdatasets with densities comparable to other datasets. Specifically, we reduced the sparsity of Fashion from $4.765\cdot 10^{-5}$ to $3.564\cdot 10^{-4}$ and Men from $6.727\cdot 10^{-5}$ to $2.860\cdot 10^{-4}$, resulting in Subfashion and Submen. We applied \RMHA, \RotatoryCon, and \None to these datasets.

Table \ref{tab:stable_best_ablation} shows that \RMHA consistently produced the best results across both subdatasets, confirming that its effectiveness is not solely due to sparsity. In Subfashion,  \HitDev increased to $15.105$, and \NDCGDev to $17.935$, while Submen experienced a small reduction in deviations (\HitDev $7.296$, \NDCGDev $9.096$). Although Men saw a slight reduction on deviation (Table \ref{tab:Deviation_Ablation}), it wasn’t substantial.

Additionally, we created Subgames for comparison, reducing the Games dataset’s sparsity to see if it influences performance. Interestingly, \RotatoryCon remained the best encoding for this more stable dataset (Table \ref{tab:stable_best_ablation}), suggesting that even after sparsity reduction, dataset characteristics like deviation may drive encoding performance.

\paragraph{Deviation Reduction:}
To rule out the correlation with the vector representation and isolate the effect of deviation, we further reduced the Men dataset to $10k$ users and $80k$ items ($10000$ users and $45129$ items after filtering). This reduced sparsity to $1.648\cdot 10^{-4}$ and helped stabilize the experiments, decreasing \HitDev to $5.198$ and \NDCGDev to $6.826$. The reduced deviation was enough to make \RotatoryCon the best-performing encoding. Although the performance is close to that of the \None case (Table \ref{tab:General_Ablation}), there was a shift in the optimal encoding.

These findings indicate that high-deviation datasets, even after sparsity reduction, benefit from the added stability of \RMHA. In contrast, low-deviation datasets like Subgames, perform better with \RotatoryCon~.

\begin{table}[htbp]
    \centering
    \captionsetup{justification=centering} 
    \resizebox{0.48\textwidth}{0.055\textheight}{
        \begin{tabular}{lccccccc}
            \toprule
            \textbf{Dataset} & \textbf{Users} & \textbf{Items}  & \textbf{Interac.} & \textbf{\HitDev} & \textbf{\NDCGDev} & \textbf{CI-length} & \textbf{Density} \\
            \midrule
            Subfashion & 3,840 & 5,000 & 16,080   & 15.105 & 17.935 & 24.175 & $3.564 \cdot 10^{-4}$ \\
            Submen     & 8,332 & 10,000 & 28,603  & 7.2925 & 9.0925 & 10.515 & $2.860 \cdot 10^{-4}$ \\
            Subgames   & 5,000 & 4,601 & 33,353   & 1.115 & 0.775   &   1.79 & $1.3341 \cdot 10^{-3}$ \\
            Submen 2   & 10,000 & 45,129 & 74,391 & 5.1975 & 6.8225 &   8.31 & $1.648 \cdot 10^{-4}$ \\
            \bottomrule
        \end{tabular}
        }
    \captionsetup{width=.48\textwidth}
    \caption{Statistics of the ablations: Deviations for  Hit@10 and NDCG without Encoding. Together with the average length of the $CI$ intervals, and the density.}
    \label{tab:Deviation_Ablation}
    \vspace{-10pt}
\end{table}

\vspace{-10pt}

\section{Conclusion}
In this study, we emphasize that encodings are not mere technical details but rather critical components that can significantly influence both the performance and stability of SRS. Our findings reveal that high-deviation datasets benefit from relative encodings, such as \RMHA, which not only stabilize training but also yield optimal results. These encodings play a vital role in mitigating performance fluctuations, allowing the model to consistently perform well across different runs. On the other hand, for low-deviation datasets, our proposed \Rotatory encoding offers a robust alternative that further improves results without introducing unnecessary complexity.

Importantly, at the time of publication, we achieved state-of-the-art performance using only these encoding methods, surpassing previous benchmarks and demonstrating greater reliability. Our results indicate that encodings are a pivotal factor in ensuring the stability of transformer-based models, especially in recommendation systems. We encourage the research community to place greater emphasis on positional encodings, as they can drastically enhance model effectiveness, particularly in tasks involving sequential data.

In the Appendix \ref{sec:decision_diagram}, there is a decision tree for the encoding election.


\bibliography{custom}
\bibliographystyle{acl_natbib}

\clearpage
\onecolumn

\appendix
\section{Appendix}
\subsection{Experiments}
\label{sec:complete_tables}
In this section, we present all our results for the four datasets, thereby completing the missing experiments from the sub-tables in the main corpus. In tables \ref{tab:beauty}, \ref{tab:fashion}, and \ref{tab:men}, we can find \LongROPEMHA, which is not presented in Table \ref{tab:encodings_table}. These are just a few cases in which we extend the number of epochs since we found that for \ROPEMHA, it takes longer to reach its maximum.
\begin{table*}[h]
\captionsetup{justification=centering} 
\footnotesize\addtolength{\tabcolsep}{-1pt}
\begin{tabular}{llrrrrrrcr}
\toprule
Act & encoding &  nmax & \HitMean & \HitDev & \NDCGMean &  \NDCGDev &  runs & CI &  CI-length \\
\midrule
leaky &\None & 0.0001 &56.00 &0.96 & 37.93 & 0.65 &9 & (55.37, 56.63) &  1.26 \\
 silu &\None & 0.0001 &57.03 &0.83 & 38.42 & 0.68 &9 & (56.49, 57.57) &  1.08 \\
leaky &\None & 0.1000 &49.02 &1.18 & 30.73 & 1.21 &9 & (48.25, 49.79) &  1.54 \\
 silu &\None & 0.1000 &49.69 &0.77 & 31.35 & 0.94 &9 & (49.19, 50.19) &  1.00 \\
leaky &\LearntCon & 0.0001 &56.36 &0.65 & 37.94 & 1.54 &3 &  (55.62, 57.1) &  1.48 \\
 silu &\LearntCon & 0.0001 &56.69 &0.51 & 37.82 & 0.65 &3 & (56.11, 57.27) &  1.16 \\
leaky &\LearntCon & 0.1000 &49.29 &0.71 & 30.98 & 0.99 &3 & (48.49, 50.09) &  1.60 \\
 silu &\LearntCon & 0.1000 &49.85 &1.18 & 31.42 & 1.48 &3 & (48.51, 51.19) &  2.68 \\
leaky &  \AbsCon & 0.0001 &67.93 &4.17 & 48.71 & 3.59 &5 & (64.27, 71.59) &  7.32 \\
 silu &  \AbsCon & 0.0001 &65.87 &5.23 & 45.61 & 5.12 &5 & (61.29, 70.45) &  9.16 \\
leaky &  \AbsCon & 0.1000 &51.42 &1.41 & 32.10 & 0.46 &3 & (49.82, 53.02) &  3.20 \\
 silu &  \AbsCon & 0.1000 &54.54 &3.33 & 35.43 & 3.34 &5 & (51.62, 57.46) &  5.84 \\
leaky &  \RotatoryCon & 0.0001 &61.16 &0.77 & 42.83 & 0.48 &3 & (60.29, 62.03) &  1.74 \\
 silu &  \RotatoryCon & 0.0001 &61.68 &1.04 & 42.31 & 0.97 &3 &  (60.5, 62.86) &  2.36 \\
leaky &  \RotatoryCon & 0.1000 &50.51 &1.59 & 32.25 & 1.76 &3 & (48.71, 52.31) &  3.60 \\
 silu &  \RotatoryCon & 0.1000 &50.88 &1.00 & 32.70 & 0.87 &3 & (49.75, 52.01) &  2.26 \\
leaky &  \Learnt & 0.0001 &59.83 &3.71 & 39.85 & 3.11 &3 & (55.63, 64.03) &  8.40 \\
 silu &  \Learnt & 0.0001 &67.62 &6.95 & 46.33 & 7.08 &5 & (61.53, 73.71) & 12.18 \\
leaky &  \Learnt & 0.1000 &74.49 &17.19 & 51.36 &22.65 &6 & (60.74, 88.24) & 27.50 \\
 silu &  \Learnt & 0.1000 &61.75 &7.84 & 40.02 & 7.34 &10 & (56.89, 66.61) &  9.72 \\
 silu & \LongRotatory & 0.0001 &61.87 &1.04 & 42.60 & 0.98 &2 & (60.43, 63.31) &  2.88 \\
 silu & \LongRotatory & 0.1000 &49.33 &2.08 & 30.89 & 2.28 &2 & (46.45, 52.21) &  5.76 \\
leaky &\Abs & 0.0001 &63.65 &8.36 & 43.68 & 8.36 &13 & (59.11, 68.19) &  9.08 \\
 silu &\Abs & 0.0001 &67.27 &10.59 & 45.34 &10.75 &15 & (61.91, 72.63) & 10.72 \\
leaky &\Abs & 0.1000 &57.25 &10.84 & 37.65 &10.73 &12 & (51.12, 63.38) & 12.26 \\
 silu &\Abs & 0.1000 &61.23 &15.24 & 39.14 &11.78 &16 &  (53.76, 68.7) & 14.94 \\
leaky &\RMHA & 0.0001 &56.29 &0.81 & 37.66 & 1.14 &3 & (55.37, 57.21) &  1.84 \\
 silu &\RMHA & 0.0001 &57.00 &0.35 & 37.22 & 0.32 &3 &   (56.6, 57.4) &  0.80 \\
leaky &\RMHA & 0.1000 &51.30 &0.32 & 33.38 & 0.15 &3 & (50.94, 51.66) &  0.72 \\
 silu &\RMHA & 0.1000 &51.35 &0.42 & 33.04 & 0.12 &3 & (50.87, 51.83) &  0.96 \\
leaky &  \RMHAROPEONE & 0.0001 &55.33 &1.22 & 37.89 & 1.24 &3 & (53.95, 56.71) &  2.76 \\
 silu &  \RMHAROPEONE & 0.0001 &57.25 &1.16 & 38.76 & 1.01 &3 & (55.94, 58.56) &  2.62 \\
leaky &  \RMHAROPEONE & 0.0001 &55.95 &0.21 & 38.41 & 0.65 &3 & (55.71, 56.19) &  0.48 \\
 silu &  \RMHAROPEONE & 0.0001 &56.23 &0.63 & 37.69 & 0.55 &3 & (55.52, 56.94) &  1.42 \\
leaky & \ROPEMHA & 0.0001 &56.30 &0.92 & 38.60 & 0.82 &3 & (55.26, 57.34) &  2.08 \\
 silu & \ROPEMHA & 0.0001 &56.59 &0.72 & 37.21 & 0.98 &3 &  (55.78, 57.4) &  1.62 \\
leaky & \ROPEMHA & 0.1000 &50.04 &0.80 & 31.91 & 0.89 &3 & (49.13, 50.95) &  1.82 \\
 silu & \ROPEMHA & 0.1000 &49.15 &0.71 & 31.31 & 1.01 &3 & (48.35, 49.95) &  1.60 \\
leaky &\Rotatory & 0.0001 &58.69 &1.31 & 40.83 & 0.71 &5 & (57.54, 59.84) &  2.30 \\
 silu &\Rotatory & 0.0001 &61.72 &1.14 & 42.54 & 1.52 &5 & (60.72, 62.72) &  2.00 \\
leaky &\Rotatory & 0.1000 &49.11 &1.17 & 30.57 & 1.28 &5 & (48.08, 50.14) &  2.06 \\
 silu &\Rotatory & 0.1000 &49.40 &0.96 & 31.14 & 1.04 &5 & (48.56, 50.24) &  1.68 \\
\bottomrule
\end{tabular}
    \caption{Beauty results with different positional encodings. Original results from \cite{CARCA} at the end.}
    \label{tab:beauty}
\end{table*}
\vspace{-45pt}
\begin{table*}[h]
\captionsetup{justification=centering} 
\footnotesize\addtolength{\tabcolsep}{-1pt}
\begin{tabular}{llrrrrrrcr}
\toprule
Act & encoding &  nmax & \HitMean & \HitDev & \NDCGMean &  \NDCGDev &  runs & CI &  CI-length \\
\midrule
leaky & \None & 0.0001 &65.44 &3.54 & 46.46 & 9.99 &9 & (63.13, 67.75) &  4.62 \\
 silu & \None & 0.0001 &70.38 &1.82 & 53.34 & 7.00 &6 & (68.92, 71.84) &  2.92 \\
leaky & \None & 0.1000 &54.04 &7.99 & 38.82 & 9.53 &3 &  (45.0, 63.08) & 18.08 \\
 silu & \None & 0.1000 &61.53 &11.84 & 48.13 &14.97 &3 & (48.13, 74.93) & 26.80 \\
leaky & \LearntCon & 0.0001 &69.31 &7.74 & 50.73 &12.02 &14 & (65.26, 73.36) &  8.10 \\
 silu & \LearntCon & 0.0001 &69.81 &2.03 & 49.49 & 7.59 &7 & (68.31, 71.31) & 3.00 \\
leaky & \LearntCon & 0.1000 &65.29 &7.61 & 52.91 & 9.70 &10 & (60.57, 70.01) & 9.44 \\
 silu & \LearntCon & 0.1000 &66.15 &7.47 & 50.59 & 9.46 &14 & (62.24, 70.06) & 7.82 \\
leaky & \AbsCon & 0.0001 &67.30 &4.48 & 52.26 & 9.94 &10 & (64.52, 70.08) &  5.56 \\
 silu & \AbsCon & 0.0001 &70.83 &1.64 & 59.93 & 1.95 &6 & (69.52, 72.14) &  2.62 \\
leaky & \AbsCon & 0.1000 &64.93 &10.58 & 52.21 &13.11 &16 & (59.75, 70.11) & 10.36 \\
 silu & \AbsCon & 0.1000 &65.14 &8.20 & 51.71 &11.06 &13 &  (60.68, 69.6) & 8.92 \\
leaky & \RotatoryCon & 0.0001 &69.13 &2.80 & 55.33 & 7.69 &9 &  (67.3, 70.96) & 3.66 \\
 silu & \RotatoryCon & 0.0001 &71.16 &1.38 & 48.61 & 8.62 &7 & (70.14, 72.18) & 2.04 \\
leaky & \RotatoryCon & 0.1000 &68.61 &2.76 & 57.08 & 3.72 &7 & (66.57, 70.65) & 4.08 \\
 silu & \RotatoryCon & 0.1000 &65.23 &7.76 & 52.99 & 9.31 &11 & (60.64, 69.82) & 9.18 \\
leaky & \Learnt & 0.0001 &68.37 &5.75 & 51.41 &12.43 &13 &  (65.24, 71.5) & 6.26 \\
 silu & \Learnt & 0.0001 &67.63 &5.75 & 46.36 &11.95 &13 &  (64.5, 70.76) & 6.26 \\
leaky & \Learnt & 0.1000 &64.99 &9.00 & 47.66 &14.48 &16 &  (60.58, 69.4) & 8.82 \\
 silu & \Learnt & 0.1000 &63.94 &9.49 & 46.44 &10.58 &16 & (59.29, 68.59) & 9.30 \\
leaky &\LongerRMHA & 0.0001 &76.20 & 0.51 & 49.85 & 3.92 & 6 & (75.79, 76.61) & 0.82\\
 silu &\LongerRMHA & 0.0001 & 77.00 & 0.48 & 49.40 & 0.54 & 6 & (76.62, 77.38) & 0.76 \\
leaky & \LongROPEMHA & 0.0001 &68.63 &3.97 & 51.05 &11.71 &5 & (65.15, 72.11) & 6.96 \\
 silu & \LongROPEMHA & 0.0001 &71.50 &1.22 & 49.69 & 6.41 &7 &   (70.6, 72.4) & 1.80 \\
leaky &\Abs & 0.0001 &66.39 &4.61 & 51.99 & 9.51 &10 & (63.53, 69.25) & 5.72 \\
 silu &\Abs & 0.0001 &69.52 &4.76 & 49.72 &10.39 &10 & (66.57, 72.47) & 5.90 \\
leaky &\Abs & 0.1000 &59.75 &9.74 & 44.73 &12.86 &15 & (54.82, 64.68) & 9.86 \\
 silu &\Abs & 0.1000 &58.39 &8.48 & 41.86 &11.31 &15 &  (54.1, 62.68) & 8.58 \\
leaky & \RMHA & 0.0001 &76.46 &0.25 & 49.49 & 2.18 &6 & (76.26, 76.66) & 0.40 \\
 silu & \RMHA & 0.0001 &77.26 &0.24 & 49.75 & 2.33 &6 & (77.07, 77.45) & 0.38 \\
leaky & \RMHA & 0.1000 &60.63 &5.14 & 40.75 & 8.06 &11 & (57.59, 63.67) & 6.08 \\
 silu & \RMHA & 0.1000 &63.61 &4.41 & 39.48 & 7.97 &11 &  (61.0, 66.22) & 5.22 \\
leaky & \RMHAROPEONE & 0.0001 &73.03 &6.52 & 62.96 & 8.59 &3 & (65.65, 80.41) & 14.76 \\
 silu & \RMHAROPEONE & 0.0001 &71.70 &1.82 & 51.07 &11.17 &3 & (69.64, 73.76) & 4.12 \\
leaky & \RMHAROPEONE & 0.0001 &69.27 &5.02 & 54.24 &11.80 &6 & (65.25, 73.29) & 8.04 \\
 silu & \RMHAROPEONE & 0.0001 &70.49 &1.66 & 49.17 &10.70 &6 & (69.16, 71.82) & 2.66 \\
leaky &\ROPEMHA & 0.0001 &66.73 &4.81 & 47.49 &12.62 &5 & (62.51, 70.95) & 8.44 \\
 silu &\ROPEMHA & 0.0001 &71.40 &1.75 & 52.83 & 8.88 &5 & (69.87, 72.93) & 3.06 \\
leaky &\ROPEMHA & 0.1000 &61.24 &8.78 & 48.13 &10.86 &6 & (54.21, 68.27) & 14.06 \\
 silu &\ROPEMHA & 0.1000 &62.79 &8.79 & 49.27 &10.69 &8 &  (56.7, 68.88) & 12.18 \\
leaky &\Rotatory & 0.0001 &68.95 &2.90 & 53.66 & 9.58 &15 & (67.48, 70.42) &  2.94 \\
 silu &\Rotatory & 0.0001 &70.87 &1.22 & 48.23 & 7.43 &15 & (70.25, 71.49) &  1.24 \\
leaky &\Rotatory & 0.1000 &62.02 &10.97 & 48.79 &13.69 &18 & (56.95, 67.09) & 10.14 \\
 silu &\Rotatory & 0.1000 &62.68 &8.53 & 49.41 &10.86 &15 &  (58.36, 67.0) &  8.64 \\
\bottomrule
leaky &  \None &  0.0001 & 59.1 & -- & 38.1 & -- & 5 & -- &  -- \\
\bottomrule
\end{tabular}
    \caption{Fashion Dataset results with different positional encodings. Results from \cite{CARCA} at the end.}
    \label{tab:fashion}
\end{table*}
\vspace{-45pt}
\begin{table*}[h]
\captionsetup{justification=centering} 
\scriptsize\addtolength{\tabcolsep}{-2pt}
\begin{tabular}{llrrrrrrcr}
\toprule
Act & encoding &  nmax & \HitMean & \HitDev & \NDCGMean &  \NDCGDev &  runs & CI &  CI-length \\
\midrule
leaky & \None & 0.0001 &65.19 &6.37 & 48.99 &11.96 &9 & (61.03, 69.35) & 8.32 \\
 silu & \None & 0.0001 &64.90 &3.06 & 47.51 & 9.84 &6 & (62.45, 67.35) & 4.90 \\
leaky & \None & 0.1000 &69.69 &1.28 & 57.94 & 1.69 &3 & (68.24, 71.14) & 2.90 \\
 silu & \None & 0.1000 &65.66 &0.89 & 53.12 & 0.67 &3 & (64.65, 66.67) & 2.02 \\
leaky & \LearntCon & 0.0001 &71.34 &7.13 & 57.92 &13.07 &7 & (66.06, 76.62) & 10.56 \\
 silu & \LearntCon & 0.0001 &66.64 &4.49 & 48.38 &12.61 &7 & (63.31, 69.97) &  6.66 \\
leaky & \LearntCon & 0.1000 &73.86 &7.18 & 58.89 & 8.15 &7 & (68.54, 79.18) & 10.64 \\
 silu & \LearntCon & 0.1000 &62.18 &7.83 & 47.82 &10.83 &7 & (56.38, 67.98) & 11.60 \\
leaky & \AbsCon & 0.0001 &70.84 &5.13 & 57.63 & 9.89 &9 & (67.49, 74.19) &  6.70 \\
 silu & \AbsCon & 0.0001 &65.55 &4.59 & 49.37 &10.61 &6 & (61.88, 69.22) &  7.34 \\
leaky & \AbsCon & 0.1000 &67.31 &6.18 & 54.49 & 9.17 &9 & (63.27, 71.35) &  8.08 \\
 silu & \AbsCon & 0.1000 &63.33 &7.50 & 48.62 &10.54 &9 & (58.43, 68.23) &  9.80 \\
leaky & \RotatoryCon & 0.0001 &70.64 &5.29 & 59.30 & 6.98 &10 & (67.36, 73.92) & 6.56 \\
 silu & \RotatoryCon & 0.0001 &67.18 &5.82 & 50.27 &13.21 &10 & (63.57, 70.79) & 7.22 \\
leaky & \RotatoryCon & 0.1000 &70.92 &3.92 & 59.84 & 4.88 &6 & (67.78, 74.06) & 6.28 \\
 silu & \RotatoryCon & 0.1000 &63.88 &7.42 & 50.69 & 9.75 &10 & (59.28, 68.48) & 9.20 \\
leaky & \Learnt & 0.0001 &72.08 &4.44 & 56.55 & 9.98 &7 & (68.79, 75.37) &  6.58 \\
 silu & \Learnt & 0.0001 &66.18 &7.81 & 46.61 &11.21 &8 & (60.77, 71.59) & 10.82 \\
leaky & \Learnt & 0.1000 &67.71 &9.94 & 49.97 & 9.10 &8 & (60.82, 74.6) & 13.78 \\
 silu & \Learnt & 0.1000 &67.15 &12.19 & 51.53 &14.44 &8 &(58.7, 75.6) & 16.90 \\
leaky & \LongerRMHA & 0.0001 &68.72 &1.47 & 43.46 & 0.79 &3 & (67.06, 70.38) & 3.32 \\
 silu & \LongerRMHA & 0.0001 &70.13 & 0.42 & 46.41 & 4.18 & 6 & (69.79, 70.47) & 0.68 \\
leaky & \LongROPEMHA & 0.0001 &68.74 &4.59 & 56.80 & 5.87 &4 & (64.24, 73.24) & 9.00 \\
 silu & \LongROPEMHA & 0.0001 &67.31 &3.54 & 51.40 & 8.62 &5 & (64.21, 70.41) & 6.20 \\
leaky & \Abs & 0.0001 &65.46 &6.64 & 50.75 &11.48 &9 & (61.12, 69.8) & 8.68 \\
 silu & \Abs & 0.0001 &64.15 &2.94 & 42.68 & 8.87 &6 & (61.8, 66.5) & 4.70 \\
leaky & \Abs & 0.1000 &69.20 &4.78 & 54.43 & 8.87 &6 & (65.38, 73.02) & 7.64 \\
 silu & \Abs & 0.1000 &65.30 &8.34 & 52.33 &10.38 &9 & (59.85, 70.75) & 10.90 \\
leaky &\RMHA & 0.0001 &68.65 &0.48 & 43.24 & 0.35 &6 & (68.27, 69.03) & 0.76 \\
 silu &\RMHA & 0.0001 &69.70 &0.64 & 43.75 & 1.44 &6 & (69.19, 70.21) & 1.02 \\
leaky &\RMHA & 0.1000 &58.72 &3.28 & 41.22 & 6.16 &6 & (56.1, 61.34) & 5.24 \\
 silu &\RMHA & 0.1000 &60.77 &1.59 & 40.89 & 5.90 &6 & (59.5, 62.04) & 2.54 \\
leaky & \RMHAROPEONE & 0.0001 &69.77 &7.86 & 54.24 &17.65 &3 & (60.88, 78.66) & 17.78 \\
 silu & \RMHAROPEONE & 0.0001 &62.14 &0.74 & 37.49 & 3.31 &3 & (61.3, 62.98) &  1.68 \\
leaky & \RMHAROPEONE & 0.0001 &67.82 &6.59 & 54.10 &12.14 &6 & (62.55, 73.09) & 10.54 \\
 silu & \RMHAROPEONE & 0.0001 &66.24 &4.18 & 49.30 &11.44 &6 & (62.9, 69.58) &  6.68 \\
leaky &\ROPEMHA & 0.0001 &65.29 &6.06 & 49.86 &11.15 &7 &  (60.8, 69.78) & 8.98 \\
 silu &\ROPEMHA & 0.0001 &63.93 &3.25 & 46.10 & 7.61 &5 & (61.08, 66.78) & 5.70 \\
leaky &\ROPEMHA & 0.1000 &62.76 &9.47 & 49.77 &12.13 &6 & (55.18, 70.34) & 15.16 \\
 silu &\ROPEMHA & 0.1000 &61.24 &8.88 & 46.52 &12.55 &6 & (54.13, 68.35) & 14.22 \\
leaky & \Rotatory & 0.0001 &67.71 &5.57 & 52.99 &11.95 &9 & (64.07, 71.35) &  7.28 \\
 silu & \Rotatory & 0.0001 &67.07 &3.63 & 54.88 & 4.54 &6 & (64.17, 69.97) &  5.80 \\
leaky & \Rotatory & 0.1000 &68.07 &4.68 & 56.31 & 6.21 &6 & (64.33, 71.81) &  7.48 \\
 silu & \Rotatory & 0.1000 &63.07 &8.92 & 49.38 &12.30 &    10 &  (57.54, 68.6) & 11.06 \\
\bottomrule
leaky &  \None &  0.0001 & 55.0 & -- & 34.9 & -- & 5 & -- &  -- \\
\bottomrule
\end{tabular}
    \caption{{\footnotesize Men Dataset results with different positional encodings. Results from \cite{CARCA} at the end.}}
    \label{tab:men}
\end{table*}
\vspace{-45pt}
\begin{table*}[h]
    \captionsetup{justification=centering} 
\scriptsize\addtolength{\tabcolsep}{-2pt}
\begin{tabular}{llrrrrrrcr}
\toprule
Act & encoding &  nmax & \HitMean & \HitDev & \NDCGMean &  \NDCGDev &  runs & CI &  CI-length \\
\midrule
 leaky & \None &   NaN & 78.95 & 1.25 &  53.37 &  1.21 & 6 & (77.95, 79.95) &   2.00 \\
  silu & \None &   NaN & 78.49 & 0.98 &  52.83 &  0.74 &10 &  (77.88, 79.1) &   1.22 \\
 leaky & \LearntCon &   NaN & 78.34 & 1.28 &  54.14 &  1.39 & 6 & (77.32, 79.36) &   2.04 \\
  silu & \LearntCon &   NaN & 79.32 & 1.50 &  53.68 &  1.97 & 6 & (78.12, 80.52) &   2.40 \\
 leaky & \RotatoryCon &   NaN & 80.62 & 0.55 &  56.07 &  0.79 & 6 & (80.18, 81.06) &   0.88 \\
  silu & \RotatoryCon &   NaN & 79.61 & 2.17 &  54.80 &  2.71 & 6 & (77.87, 81.35) &   3.48 \\
 leaky & \AbsCon &   NaN & 79.71 & 1.28 &  54.93 &  1.50 & 6 & (78.69, 80.73) &   2.04 \\
  silu & \AbsCon &   NaN & 79.34 & 1.67 &  54.23 &  1.84 &10 &  (78.3, 80.38) &   2.08 \\
 leaky & \Learnt &   NaN & 78.64 & 2.05 &  54.30 &  2.41 & 6 &  (77.0, 80.28) &   3.28 \\
  silu & \Learnt &   NaN & 79.82 & 1.88 &  54.30 &  2.95 & 8 & (78.52, 81.12) &   2.60 \\
 leaky & \LongRotatoryCon &   NaN & 80.21 & 1.51 &  55.06 &  1.93 & 6 &  (79.0, 81.42) &   2.42 \\
  silu & \LongRotatoryCon &   NaN & 80.29 & 0.47 &  55.18 &  0.55 & 6 & (79.91, 80.67) &   0.76 \\
 leaky &\LongRotatory &   NaN & 78.00 & 1.10 &  52.30 &  1.05 & 6 & (77.12, 78.88) &   1.76 \\
  silu &\LongRotatory &   NaN & 76.41 & 1.00 &  51.48 &  1.10 & 6 & (75.61, 77.21) &   1.60 \\
 leaky & \Abs &   NaN & 75.27 & 2.30 &  49.61 &  2.90 & 6 & (73.43, 77.11) &   3.68 \\
  silu & \Abs &   NaN & 73.27 & 3.25 &  47.22 &  3.75 & 6 & (70.67, 75.87) &   5.20 \\
 leaky & \RMHA &   NaN & 78.85 & 0.51 &  53.32 &  0.44 &12 & (78.56, 79.14) &   0.58 \\
  silu & \RMHA &   NaN & 76.59 & 0.78 &  51.54 &  0.43 & 6 & (75.97, 77.21) &   1.24 \\
 leaky & \RMHAROPEONE &   NaN & 78.82 & 0.62 &  53.45 &  0.42 & 6 & (78.32, 79.32) &   1.00 \\
  silu & \RMHAROPEONE &   NaN & 78.27 & 1.19 &  52.85 &  0.88 & 6 & (77.32, 79.22) &   1.90 \\
 leaky & \ROPEMHA &   NaN & 78.97 & 0.33 &  53.64 &  0.27 & 6 & (78.71, 79.23) &   0.52 \\
  silu & \ROPEMHA &   NaN & 77.56 & 1.17 &  52.04 &  1.25 & 6 &  (76.62, 78.5) &   1.88 \\
 leaky & \Rotatory &   NaN & 76.56 & 2.17 &  51.09 &  2.29 & 6 &  (74.82, 78.3) &   3.48 \\
  silu & \Rotatory &   NaN & 76.21 & 1.77 &  51.06 &  2.04 & 6 & (74.79, 77.63) &   2.84 \\
\bottomrule
leaky &  \None &  NaN & 78.2 & -- & 57.3 & -- & 5 & -- &  -- \\
\bottomrule
\end{tabular}
    \caption{{\footnotesize Games Dataset results with different positional encodings. Results from \cite{CARCA} at the end.}}
    \label{tab:games}
\end{table*}
\vspace{-45pt}

\clearpage
\subsection{Sparsity and deviation experiments}
\label{sec:complete_tables_deviation}
In this section, we present tables of the datasets created by reduction to evaluate whether the encoding type depends on the vector representation of the items, deviation, and sparsity. For these cases, we considered only three encodings: \RMHA, \Rotatory, and \None. Table \ref{tab:General_Ablation} contains results for datasets with a deviation below $12$, which we deemed invalid and have therefore moved to this appendix.

\begin{table*}[!htp]
\captionsetup{justification=centering} 
\small\addtolength{\tabcolsep}{-2pt}
\begin{tabular}{llrrrrrrcr}
\toprule
Act & encoding &  nmax & \HitMean & \HitDev & \NDCGMean &  \NDCGDev &  runs & CI &  CI-length \\
\midrule
     leaky &       \None & 0.0001 & 48.00 & 22.19 & 33.53 & 26.31 & 6 & (30.24, 65.76) & 35.52 \\
      silu &       \None & 0.0001 & 66.30 & 18.61 & 55.17 & 21.99 & 6 & (51.41, 81.19) & 29.78 \\
     leaky &       \None & 0.1000 & 32.94 &  1.89 & 16.77 &  3.00 & 6 & (31.43, 34.45) &  3.02 \\
      silu &       \None & 0.1000 & 63.85 & 17.73 & 52.02 & 20.44 & 6 & (49.66, 78.04) & 28.38 \\
     leaky & \RotatoryCon& 0.0001 & 50.04 & 19.62 & 35.82 & 22.95 & 6 & (34.34, 65.74) & 31.40 \\
      silu & \RotatoryCon& 0.0001 & 63.87 & 19.99 & 52.59 & 22.77 & 6 & (47.87, 79.87) & 32.00 \\
     leaky & \RotatoryCon& 0.1000 & 49.17 & 21.02 & 35.89 & 25.10 & 6 & (32.35, 65.99) & 33.64 \\
      silu & \RotatoryCon& 0.1000 & 67.94 & 19.26 & 57.24 & 23.51 & 6 & (52.53, 83.35) & 30.82 \\
     leaky &       \RMHA & 0.0001 & 79.72 &  2.85 & 71.57 &  3.66 & 6 &  (77.44, 82.0) &  4.56 \\
      silu &       \RMHA & 0.0001 & 46.55 & 14.12 & 31.68 & 16.59 & 6 & (35.25, 57.85) & 22.60 \\
     leaky &       \RMHA & 0.1000 & 57.99 & 20.07 & 45.36 & 24.01 & 6 & (41.93, 74.05) & 32.12 \\
      silu &       \RMHA & 0.1000 & 61.14 & 14.64 & 48.46 & 18.10 & 6 & (49.43, 72.85) & 23.42 \\
\bottomrule
\end{tabular}
    \caption{SubFashion Dataset results for the best positional encodings and the None encoding case.}
    \label{tab:subfashion}
\end{table*}
\vspace{-15pt}
\begin{table*}[!htp]
\captionsetup{justification=centering} 
\small\addtolength{\tabcolsep}{-2pt}
\begin{tabular}{llrrrrrrcr}
\toprule
Act & encoding &  nmax & \HitMean & \HitDev & \NDCGMean &  \NDCGDev &  runs & CI &  CI-length \\
\midrule
     leaky & \None        & 0.0001 & 77.10 & 3.72 & 67.89 & 5.25 & 12 &   (75.0, 79.2) & 4.20 \\
      silu & \None        & 0.0001 & 76.33 & 6.11 & 67.00 & 7.08 & 12 & (72.87, 79.79) & 6.92 \\
     leaky & \None        & 0.1000 & 69.29 &15.21 & 58.64 &18.19 & 6  & (57.12, 81.46) & 24.34 \\
      silu & \None        & 0.1000 & 74.66 & 4.13 & 64.38 & 5.85 & 6  & (71.36, 77.96) & 6.60 \\
     leaky & \RotatoryCon & 0.0001 & 71.82 &11.90 & 61.69 &14.11 & 12 & (65.09, 78.55) & 13.46 \\
      silu & \RotatoryCon & 0.0001 & 75.28 & 3.63 & 65.31 & 5.03 & 12 & (73.23, 77.33) & 4.10 \\
     leaky & \RotatoryCon & 0.1000 & 73.93 & 5.87 & 63.77 & 8.00 & 6  & (69.23, 78.63) & 9.40 \\
      silu & \RotatoryCon & 0.1000 & 73.11 & 5.43 & 62.64 & 6.80 & 6  & (68.77, 77.45) & 8.68 \\
     leaky & \RMHA        & 0.0001 & 75.96 & 2.71 & 66.03 & 4.00 & 8  & (74.08, 77.84) & 3.76 \\
      silu & \RMHA        & 0.0001 & 70.44 & 9.61 & 59.67 &11.50 & 8  &  (63.78, 77.1) & 13.32 \\
     leaky & \RMHA        & 0.1000 & 77.28 & 2.66 & 68.45 & 3.65 & 6  & (75.15, 79.41) & 4.26 \\
      silu & \RMHA        & 0.1000 & 72.54 & 4.57 & 61.75 & 5.96 & 6  &  (68.88, 76.2) & 7.32 \\
\bottomrule
\end{tabular}
    \caption{SubMen Dataset results for the best positional encodings and the None encoding case.}
    \label{tab:submen}
\end{table*}
\vspace{-15pt}
\begin{table*}[!htp]
\captionsetup{justification=centering} 
\small\addtolength{\tabcolsep}{-2pt}
\begin{tabular}{llrrrrrrcr}
\toprule
Act & encoding &  nmax & \HitMean & \HitDev & \NDCGMean &  \NDCGDev &  runs & CI &  CI-length \\
\midrule
     leaky &        \None &   NaN &     49.53 &     1.01 &      28.84 &      0.69 &     6 & (48.72, 50.34) &       1.62 \\
      silu &        \None &   NaN &     48.31 &     1.22 &      28.36 &      0.86 &     6 & (47.33, 49.29) &       1.96 \\
     leaky & \RotatoryCon &   NaN &     52.23 &     1.57 &      31.51 &      0.96 &     6 & (50.97, 53.49) &       2.52 \\
      silu & \RotatoryCon &   NaN &     50.73 &     0.89 &      30.48 &      0.72 &     6 & (50.02, 51.44) &       1.42 \\
     leaky &        \RMHA &   NaN &     50.43 &     0.95 &      29.96 &      0.87 &     6 & (49.67, 51.19) &       1.52 \\
      silu &        \RMHA &   NaN &     48.97 &     1.35 &      28.57 &      0.79 &     6 & (47.89, 50.05) &       2.16 \\
\bottomrule
\end{tabular}
    \captionsetup{width=.75\textwidth}
    \caption{Subgames Dataset results for the best positional encodings (\RotatoryCon and \RMHA) and the \None encoding case.}
    \label{tab:subgames}
\end{table*}
\vspace{-15pt}
\begin{table*}[!htp]
\captionsetup{justification=centering} 
\small\addtolength{\tabcolsep}{-2pt}
\begin{tabular}{llrrrrrrcr}
\toprule
Act & encoding & nmax & \HitMean & \HitDev & \NDCGMean & \NDCGDev & runs & CI &  CI-length \\
\midrule
     leaky &        \None & 0.0001 &     72.08 &     6.04 &      61.47 &      7.96 &     6 & (67.25, 76.91) &       9.66 \\
      silu &        \None & 0.0001 &     73.95 &     3.33 &      64.05 &      4.71 &     6 & (71.29, 76.61) &       5.32 \\
     leaky &        \None & 0.1000 &     71.56 &     5.38 &      60.96 &      6.93 &     6 & (67.26, 75.86) &       8.60 \\
      silu &        \None & 0.1000 &     70.51 &     6.04 &      59.29 &      7.69 &     6 & (65.68, 75.34) &       9.66 \\
     leaky & \RotatoryCon & 0.0001 &     71.52 &     6.39 &      60.71 &      8.17 &     6 & (66.41, 76.63) &      10.22 \\
      silu & \RotatoryCon & 0.0001 &     72.77 &     2.96 &      62.56 &      4.31 &     6 &  (70.4, 75.14) &       4.74 \\
     leaky & \RotatoryCon & 0.1000 &     73.12 &     2.38 &      62.86 &      2.99 &     6 & (71.22, 75.02) &       3.80 \\
      silu & \RotatoryCon & 0.1000 &     72.03 &     3.90 &      61.33 &      4.76 &     6 & (68.91, 75.15) &       6.24 \\
      silu &        \RMHA & 0.0001 &     53.67 &     8.81 &      36.15 &     12.73 &    12 & (48.69, 58.65) &       9.96 \\
     leaky &        \RMHA & 0.1000 &     61.82 &    12.74 &      48.93 &     14.82 &     6 & (51.63, 72.01) &      20.38 \\
      silu &        \RMHA & 0.1000 &     61.27 &     5.87 &      47.24 &      6.96 &     6 & (56.57, 65.97) &       9.40 \\
\bottomrule
\end{tabular}
    \caption{SubMen2 Dataset results for the best positional encodings and the None encoding case.}
    \label{tab:submen3}
\end{table*}
\vspace{-15pt}
\begin{table*}[h]
    \centering
    \captionsetup{justification=centering} 
    \resizebox{0.76\textwidth}{0.13\textheight}{
    \begin{tabular}{lllccccc}
        \toprule
        Dataset & Act & encoding & nmax & \HitMean & \HitDev & \NDCGMean & \NDCGDev \\
        \midrule
        Submen & leaky & \RMHA & 0.1000 & 77.28 & 2.66 & 68.45 & 3.65 \\
        Submen & leaky & \None & 0.0001 & 77.10 & 3.72 & 67.89 & 5.25 \\
        Submen & silu  & \None & 0.0001 & 76.33 & 6.11 & 67.00 & 7.08 \\
        Submen & leaky & \RMHA & 0.0001 & 75.96 & 2.71 & 66.03 & 4.00 \\
        \midrule
        Subfashion & leaky & \RMHA & 0.0001 & 79.72 & 2.85 & 71.57 & 3.66 \\
        Subfashion & leaky & \None & 0.1000 & 32.94 & 1.89 & 16.77 & 3.00 \\
        \midrule
        Subgames & leaky & \RotatoryCon & NaN & 52.23 & 1.57 & 31.51 & 0.96 \\
        Subgames & silu  & \RotatoryCon & NaN & 50.73 & 0.89 & 30.48 & 0.72 \\
        Subgames & leaky &        \RMHA & NaN & 50.43 & 0.95 & 29.96 & 0.87 \\
        Subgames & leaky &        \None & NaN & 49.53 & 1.01 & 28.84 & 0.69 \\
        \midrule
        Submen3 &  silu &        \None & 0.0001 & 73.95 & 3.33 & 64.05 & 4.71 \\
        Submen3 & leaky & \RotatoryCon & 0.1000 & 73.12 & 2.38 & 62.86 & 2.99 \\
        Submen3 &  silu & \RotatoryCon & 0.0001 & 72.77 & 2.96 & 62.56 & 4.31 \\
        Submen3 & leaky &        \None & 0.0001 & 72.08 & 6.04 & 61.47 & 7.96 \\
        \bottomrule
        \end{tabular}}
    \caption{The table displays the top $4$ results for each dataset on cases with a standard deviation for HIT below $12$.}
    \label{tab:General_Ablation}
\end{table*}

\vspace{-5pt}

\subsection{Resources}
\label{sec:resources}
Due to the inherent instability of these models, it was necessary to run multiple seeds to obtain reliable results. Initially, we executed each model configuration three times. However, as we analyzed the results, we determined that additional seeds were needed for cases with high-deviation, aiming to maintain a confidence interval within ten points.

All experiments were conducted on $V100$ GPUs with $32$ GB of memory. Training times varied depending on the dataset and encoding type. For the Men and Fashion datasets, training durations ranged from $4$ to $6$ hours, while for Beauty, the training times extended from $22$ to $26$ hours. Given the available resources, these experiments were spread over a period of six months.

We used the following seed values to ensure diversity in our training runs: $42$, $43$, $44$, $45$, $46$, $1809$, $1810$, $1811$, $1812$, $1741$, $1742$, $1743$, $1744$, $1745$, $123$, $234$, $345$, $456$, $567$, $678$, $789$, $890,102$, $203$, $304$, $405$, $506$, $607$, $708$, $808$, $901$, $1001$, $987$, $8765$, $7654$, and $6543$.

\subsection{Hyperparameters}
\label{sec:Hyperparameters}
In this section, we outline the hyperparameters used during training. Consistent with the approach used in CARCA \cite{CARCA}, we adopted the same baseline hyperparameters across most experiments to maintain comparability. In specific cases, we experimented with variations in the upper bound of the norm for the encodings, activation functions, as well as adjusting the training duration to explore its impact on stability and performance.

\begin{table}[h]
\centering
\begin{tabular}{|l|cccc|}
\hline
\textbf{Hyperparameter} & \textbf{Men} & \textbf{Fashion} & \textbf{Games}  & \textbf{Beauty} \\ \hline
Learning Rate & 0.000006 & 0.00001 & 0.0001 & 0.0001 \\ \hline
Max Seq. Length & 35 & 35 & 50 & 75 \\ \hline
Number of attention blocks & 3 & 3 & 3 & 3 \\ \hline
Number of attention heads & 3 & 3 & 3 & 1 \\ \hline
Dropout Rate & 0.3 & 0.3 & 0.5 & 0.5 \\ \hline
L2 reg. weight & 0.0001 & 0.0001 & 0 & 0.0001 \\ \hline
Embedding Dimension d & 390 & 390 & 90 & 90 \\ \hline
Embedding Dimension g & 1950 & 1950 & 450 & 450 \\ \hline
Residual connection in the cross-attention block & No & No & Yes & Yes \\ \hline
\end{tabular}%
\captionsetup{width=.68\textwidth}
\caption{Hyperparameters with corresponding options were used in the original CARCA model. This parameter were written in tensor flow. In "$L2$ reg. weight", the value $0$ corresponds to no bound, which in Pytorch is $Nan$, and we denote it as \None.}
\label{tab:hyperparameters}
\end{table}


\subsection{Deviations over all the trainings}
\label{sec:deviations}
In this section, we present the complete table of deviation values from all training sessions. In the main text, we focused on the training runs where the default upper bound values ($0.0001$) were used, along with $None$ for the Games dataset. The trends discussed in Section \ref{sec:Stability} remain consistent here. Encodings that incorporate concatenation generally demonstrate greater stability, with \RMHA continuing to be the most stable overall. Additionally, we observe that with extended training durations, the \Rotatory encodings surpass \ROPEMHA in stability.

\begin{table}[H]
    \centering
\begin{tabular}{l|rrrr}
\toprule
encoding & \AvgDevHit &  \AvgDevNDCG & runs &  CI-length \\ 
\midrule
\Abs             & 7.20 & 9.42 & 10.57 & 8.42 \\
\AbsCon          & 4.67 & 6.58 & 8.00 & 6.36 \\
\Learnt          & 7.43 & 9.98 & 9.07 & 10.17 \\
\LearntCon       & 4.09 & 6.53 & 6.93 & 5.66 \\
\None            & 3.05 & 5.08 & 6.71 & 5.62 \\
\RMHA            & 1.37 & 2.64 & 6.29 & 1.98 \\
\ROPEMHA         & 4.03 & 6.55 & 5.14 & 6.52 \\
\Rotatory        & 3.92 & 6.10 & 9.00 & 4.92 \\
\RotatoryCon     & 3.16 & 5.12 & 6.71 & 4.47 \\
\midrule
\RMHAROPEONE     & 2.82 & 6.54 & 4.29 & 5.46 \\
\LongROPEMHA     & 3.33 & 8.15 & 5.25 & 5.99 \\
\LongRotatory    & 1.31 & 1.35 & 4.00 & 3.00 \\
\LongRotatoryCon & 0.99 & 1.24 & 6.00 & 1.59 \\
\LongerRMHA      & 0.71 & 4.00 & 6.00 & 1.14 \\
\bottomrule
\end{tabular}
\captionsetup{width=.57\textwidth}
    \caption{Deviations among the different encodings. Those per head, like \RMHA or \ROPEMHA, are more stable than others, together with \Rotatory.}
    \label{tab:deviations_encoding_all}
\end{table}

\subsection{Losses}
\label{sec:losses}
This section shows some loss cases from our trainings showing the stability of \RMHA. The first image shows four test losses for $0.0001$ as upper bound, with silu activation. The first row corresponds to the \None encoding, while the second row is the \RMHA encoding. Each row contains the losses of four different seeds. The seeds are not paired vertically but randomly selected.

The second image shows the validation Losses for the analogous case in the Men Dataset.

\begin{figure*}[h]
\includegraphics[scale=0.5]{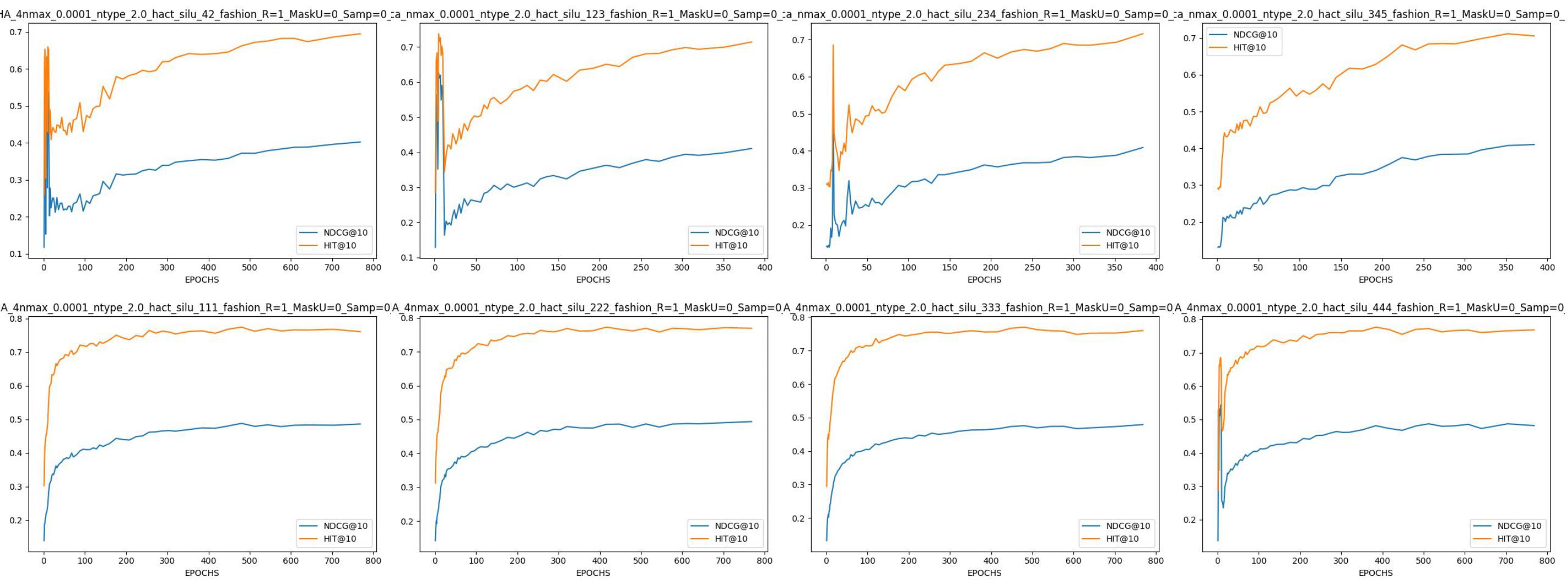}
\caption{Losses randomly selected from \None, first row, and \RMHA, second row, for the test set with $0.0001$ and silu. Dataset: Fashion}
\label{fig:fashion_test_loss1}
\end{figure*}

\begin{figure*}[h]
\centering
\includegraphics[scale=0.5]{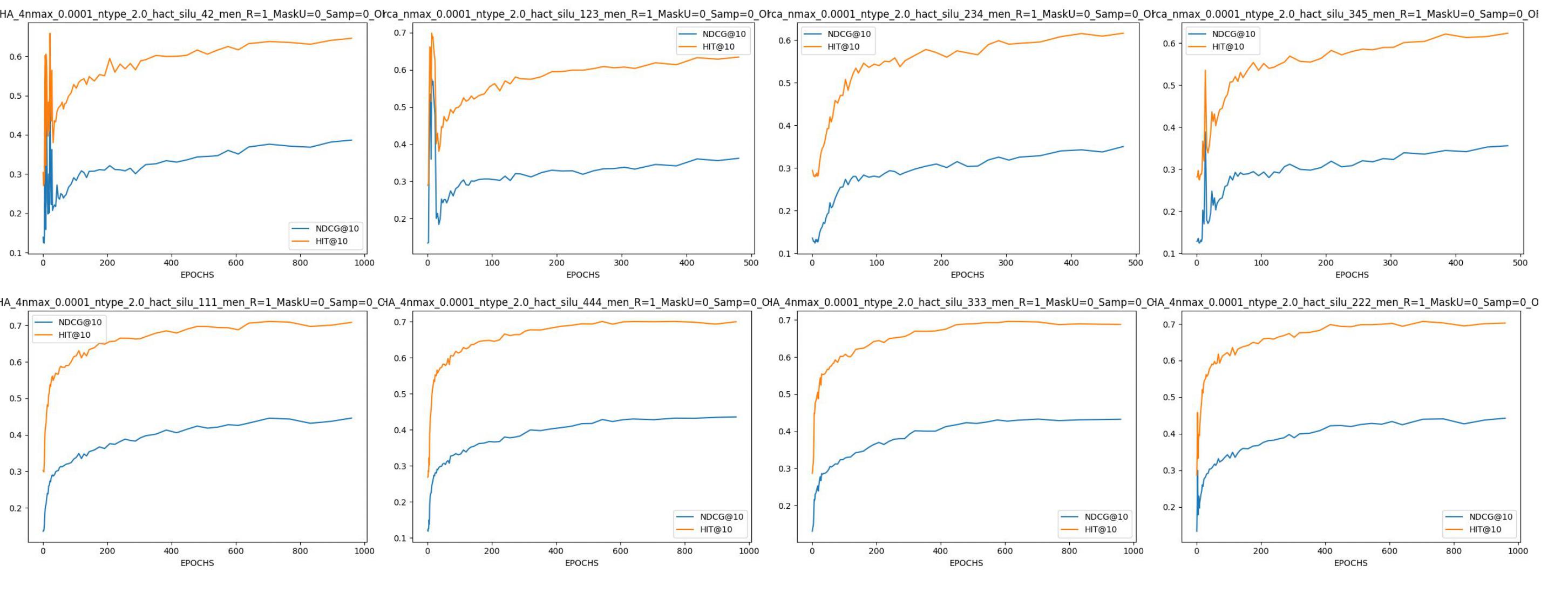}
\caption{Losses randomly selected from \None, first row, and \RMHA, second row, for the test set with $0.0001$ and silu. Dataset: Men}
\label{fig:fashion_test_loss2}
\end{figure*}

\clearpage
\subsection{Decision diagram}
\label{sec:decision_diagram}

\begin{figure*}[h]
    \centering
    \includegraphics[scale=0.8]{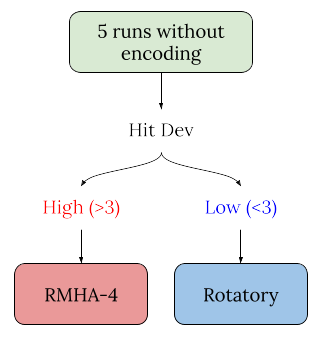}
    \caption{Decision tree for encoding selection based on dataset deviation. The diagram outlines the recommended encodings.}
    \label{fig:diagram}
\end{figure*}

\subsection{Extra experiments with SASRec++}
\label{sec:SASRec}
All of our experiments were conducted using the CARCA model \cite{CARCA}. To assess whether the encoding behaviors observed in CARCA extend to other architectures, we also evaluated the SASRec++ model, an extension of the original SASRec model \cite{SASRec}. SASRec$++$ is conceptually similar to CARCA but differs in that it replaces the transformer decoder with a simple dot product operation.

We applied SASRec++ to the Submen dataset using the following encodings: \None\!, \LearntCon\!, \RotatoryCon\!, \AbsCon\!, \Learnt\!, \Abs\!, \RMHA\!, \RMHAROPEONE\!, \ROPEMHA\!, and \Rotatory\!. As shown in Table \ref{tab:SASRec_3}, we find that \RMHA continues to outperform other encodings, maintaining its position as the best-performing encoding across all experiments. Most of the results cluster around $40\%$, with minimal differences observed between encodings. The difference between \RMHAROPEONE and other encodings is particularly small, with only slight improvements seen for \RMHA with a low \HitDev~ and \LearntCon (as seen in Table \ref{tab:SASRec_12}) with a higher \HitDev.

\begin{table*}[ht]
    \centering
    \captionsetup{justification=centering} 
    \resizebox{0.6\textwidth}{0.06\textheight}{
    \begin{tabular}{lllccccc}
    \toprule
    Dataset & Act & encoding &  nmax & \HitMean & \HitDev & \NDCGMean &  \NDCGDev \\
    \midrule
    Submen & silu & \RMHA & 0.1000 & 44.12 & 2.58 &  19.73 & 1.17 \\
    Submen & silu & \RMHA & 0.0001 & 42.45 & 2.91 &  18.91 & 1.11 \\
    Submen & leaky & \RMHAROPEONE & 0.0001 & 40.69 & 0.75 & 18.90 & 0.60 \\
    Submen & leaky & \RMHAROPEONE & 0.1000 & 40.66 & 0.74 & 18.67 & 0.39 \\
    \bottomrule
    \end{tabular}
    }
    \caption{The table displays the top $4$ results for Submen for SASRec++ on cases with a standard deviation for \HitMean below $3$.}
    \label{tab:SASRec_3}
\end{table*}

\vspace{-15pt}
\begin{table*}[ht]
    \centering
    \captionsetup{justification=centering} 
    \resizebox{0.6\textwidth}{0.06\textheight}{
    \begin{tabular}{lllccccc}
    \toprule
    Dataset & Act & encoding &  nmax & \HitMean & \HitDev & \NDCGMean &  \NDCGDev \\
    \midrule
        Submen & silu & \LearntCon & 0.0001 & 58.08 & 11.49 & 26.41 & 6.37 \\
        Submen & silu & \RMHA & 0.1000 & 44.12 & 2.58 & 19.73 & 1.17 \\
        Submen & leaky & \RMHA & 0.0001 & 42.86 & 5.52 & 20.57 & 5.34 \\
        Submen & silu & \RMHA & 0.0001 & 42.45 & 2.91 & 18.91 & 1.11 \\
    \bottomrule
    \end{tabular}
    }
    \caption{The table displays the top $4$ results for Submen for SASRec++ on cases with a standard deviation for \HitMean below $12$.}
    \label{tab:SASRec_12}
\end{table*}

\vspace{-15pt}

\end{document}